\documentclass[11pt]{amsart}
\usepackage{amssymb}
\usepackage{amsmath}
\usepackage{amsmath}
\usepackage{graphicx}
\usepackage{amsfonts}
\usepackage{amssymb}
\usepackage{stmaryrd}
\usepackage{amscd}
\numberwithin{equation}{section}

\def\cb{{\mathcal B}}
\def\cc{{\mathcal C}}

\def\bn{{\mathbb N}}

\def\br{{\mathbb R}}

\def\bz{{\mathbb Z}}

\newcommand{\dsf}{\displaystyle\frac}
\def\s{\sigma}
\def\x{\xi}
\def\k{\kappa}

\def\l{\lambda}
\def\g{\gamma}
\def\f{\varphi}
\def\z{\eta}

\def\d{\delta}

\def\O{\Omega}

\def\R{\mathbb{R}}
\def\b{\beta}
\def\m{\mu}

\def\G{\Gamma}

\newtheorem{thm}{Theorem}[section]

\newtheorem{cor}[thm]{Corollary}
\newtheorem{prop}[thm]{Proposition}

\begin{document}

\title[on Potts model with competing interactions]
{on the three state Potts model with competing interactions on the
Bethe lattice}

\author{Nasir Ganikhodjaev}
\address{Nasir Ganikhodjaev\\
Faculty of Science\\
IIUM, 53100 Kuala Lumpur, Malaysia \
and Department of Mechanics and Mathematics, NUUz\\
Vuzgorodok, 700174, Tashkent, Uzbekistan\\} \email{{\tt
nasirgani@yandex.ru}}

\author{Farrukh Mukhamedov}
\address{Farrukh Mukhemedov\\
Departamento de Fisica \\
Universidade de Aveiro\\
Campus Universit\'{a}rio de Santiago\\
3810-193 Aveiro, Portugal} \email{{\tt far75m@yandex.ru;
farruh@fis.ua.pt}}

\author{Jos\'{e} F.F. Mendes}
\address{Jos\'{e} F.F. Mendes\\
Departamento de Fisica \\
Universidade de Aveiro\\
Campus Universit\'{a}rio de Santiago\\
3810-193 Aveiro, Portugal} \email{{\tt jfmendes@fis.ua.pt}}

\maketitle

\begin{abstract}
In the present paper the three state Potts model with competing
binary interactions (with couplings $J$ and $J_p$) on the second
order Bethe lattice is considered. The recurrent equations for the
partition functions are derived.  When $J_p=0$, by means of a
construction of a special class of limiting Gibbs measures, it is
shown how these equations are related with the surface energy of
the Hamiltonian. This relation reduces the problem of describing
the limit Gibbs measures to find of solutions of a nonlinear
functional equation. Moreover, the set of ground states of the
one-level model is completely described. Using this fact, one
finds Gibbs measures (pure phases) associated with the
translation-invariant ground states. The critical temperature is
exactly found and the phase diagram is presented. The free
energies corresponding to translations-invariant Gibbs measures
are found. Certain physical quantities are calculated as well. 
\end{abstract}

\section{Introduction}

The Potts models describe a special and easily defined class of
statistical mechanics models. Nevertheless, they are richly
structured enough to illustrate almost every conceivable nuance of
the subject. In particular, they are at the center of the most
recent explosion of interest generated by the confluence of
conformal field theory,percolation theory, knot theory, quantum
groups and integrable systems. The Potts model \cite{Po} was
introduced as a  generalization of the Ising model to more than
two components. At present the Potts model encompasses a number of
problems in statistical physics (see, e.g. \cite{W}). Some exact
results about certain properties of the model were known, but more
of them are based on approximation methods. Note that there does
not exist analytical solutions on standard lattices. But
investigations of phase transitions of spin models on hierarchical
lattices showed that they make the exact calculation of various
physical quantities \cite{DGM},\cite{P1,P2},\cite{T}. Such studies
on the hierarchical lattices begun with development of the
Migdal-Kadanoff renormalization group method where the lattices
emerged as approximants of the ordinary crystal ones. On the other
hand, the study of exactly solved models deserves some general
interest in statistical mechanics \cite{Ba}. Moreover, nowadays
the investigations of statistical mechanics on non-amenable graphs
is a modern growing topic (\cite{L}). For example, Bethe lattices
are most simple hierarchical lattices with {\it non-amenable}
graph structure. This means that the ratio of the number of
boundary sites to the number of interior sites of the Bethe
lattice tends to a nonzero constant in the thermodynamic limit of
a large system, i.e. the ratio ${W_n}/{V_n}$ (see for the
definitions Sec. 2) tends to $(k-1)/(k+1)$ as $n\to\infty$, here
$k$ is the order of the lattice. Nevertheless, that the Bethe
lattice is not a realistic lattice, however, its amazing topology
makes the exact calculation of various quantities possible
\cite{L}. It is believed that several among its interesting
thermal properties could persist for regular lattices, for which
the exact calculation is far intractable. In \cite{PLM1,PLM2} the
phase diagrams of the $q$-state Potts models on the Bethe lattices
were studied and the pure phases of the the ferromagnetic Potts
model were found. In \cite{G} using those results, uncountable
number of the pure phase of the 3-state Potts model were
constructed. These investigations were based on a
measure-theoretic approach developed in
\cite{Ge},\cite{Pr},\cite{S},\cite{P1,P2}. The Bethe lattices were
fruitfully used to have a deeper insight into the behavior of the
Potts models. The structure of the Gibbs measures of the Potts
models has been investigated in \cite{G,GR}. Certain algebraic
properties of the Gibbs measures associated with the model have
been considered in \cite{M}.

It is known that the Ising model with competing interactions was
originally considered by Elliot \cite{E} in order to describe
modulated structures in rare-earth systems. In \cite{BB} the
interest to the model was renewed and studied by means of an
iteration procedure. The Ising type models on the Bethe lattices
with competing interactions appeared in a pioneering work
Vannimenus \cite{V}, in which the physical motivations for the
urgency of the study such models were presented. In \cite{YOS,TY}
the infinite-coordination limit of the model introduced by
Vannimenus was considered. It was also found a phase diagram which
was similar to that model studied in \cite{BB}. In
\cite{MTA},\cite{SC} other generalizations of the model were
studied. In all of those works the phase diagrams of such models
were found numerically, so there were not exact solutions of the
phase transition problem. Note that the ordinary Ising model on
Bethe lattices was investigated in \cite{BG,BRZ1,BRZ2,BRSSZ},
where such model was rigourously investigated. In
\cite{GPW1,GPW2},\cite{MR1,MR2} the Ising model with competing
interactions has been rigourously studied, namely for this model a
phase transition problem was exactly solved and a critical curve
was found as well. For such a model it was shown that a phase
transition occurs for the medium temperature values, which
essentially differs from the well-known results for the ordinary
Ising model, in which a phase transition occurs at low
temperature. Moreover, the structure of the set of periodic Gibbs
measures was described. While studying such models the appearance
of nontrivial magnetic orderings were discovered.

Since the Ising model corresponds to the two-state Potts model,
therefore it is naturally to consider $q$-state Potts model with
competing interactions on the Bethe lattices. Note that such kind
of models were studied in \cite{NS},\cite{Ma},\cite{Mo1,Mo2} on
standard $\bz^d$ and other lattices. In the present paper we are
going to study a phase transition problem for the three-state
ferromagnetic Potts model with competing interactions on a Bethe
lattice of order two. In this paper we will use a
measure-theoretic approach developed in \cite{Ge,S}, which enables
us to solve exactly such a model.

The paper is organized as follows. In section 2 we give some
preliminary definitions of the model with competing ternary (with
couplings $J$ and $J_p$) and binary interactions on a Bethe
lattice. In section 3 we derive recurrent equations for the
partition functions. To show how the derived recurrent equations
are related with the surface energy of the  Hamiltonian, we give a
construction of a special class of limiting Gibbs measures for the
model at $J_p=0$. Moreover, the problem of describing the limit
Gibbs measures is reduced to a problem of solving a nonlinear
functional equation. In section 4 the set of ground states of the
model is completely described. Using this fact and the recurrent
equations, in section 5, one finds Gibbs measures (pure phases)
associated with the translation-invariant ground states. A curve
of the critical temperature is exactly found, under one there
occurs a phase transition. In section 6, we prove the existence of
the free energy. The free energy of the translations-invariant
Gibbs measures is also calculated. Some physical quantities are
computed as well. Discussions of the results are given in the last
section.

\section{Preliminaries}

Recall that the Bethe lattice $\Gamma^k$ of order $ k\geq 1 $ is
an infinite tree, i.e., a graph without cycles, such that from
each vertex of which issues exactly $ k+1 $ edges. Let
$\Gamma^k=(V, \Lambda),$ where $V$ is the set of vertices of $
\Gamma^k$, $\Lambda$ is the set of edges of $ \Gamma^k$. Two
vertices $x$ and $y$ are called {\it nearest neighbors} if there
exists an edge $l\in\Lambda$ connecting them, which is denoted by
$l=<x,y>$. A collection of the pairs $<x,x_1>,...,<x_{d-1},y>$ is
called a {\it path} from $x$ to $y$. Then the distance $d(x,y),
x,y\in V$, on the Bethe lattice, is the number of edges in the
shortest path from $x$ to $y$.

For a fixed $x^0\in V$ we set
\begin{equation*}
W_n=\{x\in V| d(x,x^0)=n\}, \ \ \ V_n=\cup_{m=1}^n W_m,
\end{equation*}
\begin{equation*}
L_n=\{l=<x,y>\in L | x,y\in V_n\}. \end{equation*}
Denote
$$
S(x)=\{y\in W_{n+1} :  d(x,y)=1 \}, \ \ x\in W_n, $$ this set is
called a set of {\it direct successors} of $x$.

For the sake of simplicity we put $|x|=d(x,x^0)$, $x\in V$. Two
vertices $x,y\in V$ are called {\it the second neighbors} if
$d(x,y)=2$. Two vertices $x,y\in V$ are called {\it one level
next-nearest-neighbor vertices} if there is a vertex $z\in V$ such
that  $x,y\in S(z)$, and they are denoted by $>x,y<$. In this case
the vertices $x,z,y$ are called {\it ternary} and denoted by
$<x,z,y>$. In fact, if $x$ and $y$ are one level
next-nearest-neighbor  vertices, then they are the second
neighbors with $|x|=|y|$. Therefore, we say that two second
neighbor vertices $x$ and $y$ are {\it prolonged vertices} if
$|x|\neq |y|$ and denote them by $\widetilde{>x,y<}$.

In the sequel we will consider semi-infinite Bethe lattice
$\G^2_+$ of order 2, i.e. an infinite graph without cycles with 3
edges issuing from each vertex except for $x^0$ that has only 2
edges.

\begin{figure}
\begin{center}
\includegraphics[width=10.07cm]{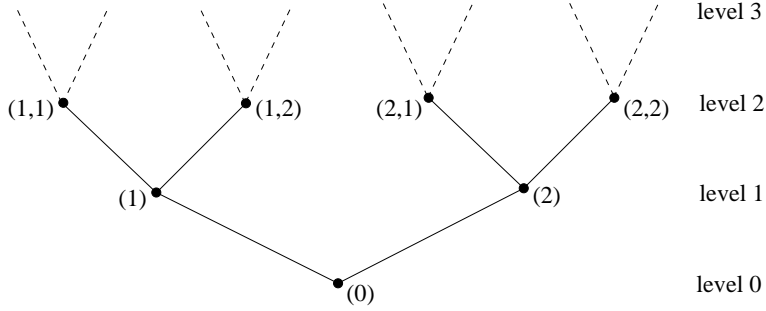}
\end{center}
\caption{The first levels of $\G_+^2$} \label{fig1}
\end{figure}

Now we are going to introduce a semigroup structure in $\G^2_+$
(see \cite{FNW}). Every vertex $x$ (except for $x^0$) of $\G^2_+$
has coordinates $(i_1,\dots,i_n)$, here $i_k\in\{1,2\}$, $1\leq
k\leq n$ and for the vertex $x^0$ we put $(0)$.  Namely, the
symbol $(0)$ constitutes level 0 and the sites $(i_1,\dots,i_n)$
form level $n$ of the lattice, i.e. for $x\in \G^2_+$,
$x=(i_1,\dots,i_n)$ we have $|x|=n$ (see Fig. \ref{fig1}).

Let us define on $\G^2_+$ a binary operation
$\circ:\G^2_+\times\G^2_+\to\G^2_+$ as follows: for any two
elements $x=(i_1,\dots,i_n)$ and $y=(j_1,\dots,j_m)$ put
\begin{equation}\label{binar1}
x\circ
y=(i_1,\dots,i_n)\circ(j_1,\dots,j_m)=(i_1,\dots,i_n,j_1,\dots,j_m)
\end{equation}
and
\begin{equation}\label{binar2}
x\circ x^0=x^0\circ x= (i_1,\dots,i_n)\circ(0)=(i_1,\dots,i_n).
\end{equation}

By means of the defined operation $\G^2_+$ becomes a
noncommutative semigroup with a unit. Using this semigroup
structure one defines translations $\tau_g:\G^2_+\to \G^2_+$,
$g\in \G^2_+$ by
\begin{equation}\label{trans1}
\tau_g(x)=g\circ x.
\end{equation}
It is clear that $\tau_{(0)}=id$.

Let $\g$ be a permutation of $\{1,2\}$. Define
$\pi^{(\g)}_{(0)}:\G^2_+\to \G^2_+$ by
\begin{equation}\label{trans2}
\left\{
\begin{array}{ll}
\pi^{(\g)}_{(0)}(0)=(0) \\[2mm]
\pi^{(\g)}_{(0)}(i_1,\dots,i_n)=(\g(i_1),\dots,i_n)\\
\end{array}
\right.
\end{equation}
for all $n\geq 1$. For any $g\in \G^2_+$ ($g\neq x^0$) define a
rotation $\pi^{(\g)}_{g}:\G^2_+\to \G^2_+$ by
\begin{equation}\label{trans3}
\pi^{(\g)}_{g}(x)=\tau_g(\pi^{(\g)}_{(0)}(x)), \ \ x\in \G^2_+.
\end{equation}

Let $G\subset \G^2_+$ be a sub-semigroup of $\G^2_+$ and
$h:\G^2_+\to \br$ be a function defined on $\G^2_+$. We say that
$h$ is {\it $G$-periodic} if $h(\tau_g(x))=h(x)$ for all $g\in G$
and $x\in \G^2_+$. Any $\G^2_+$-periodic function is called {\it
translation invariant}.  We say that $h$ is {\it quasi
$G$-periodic} if for every $g\in G$ one holds
$h(\pi^{(\g)}_{g}(x))=h(x)$ for all $x\in \G^2_+$ except for a
finite number of elements of $\G^2_+$.

Put
\begin{equation}\label{sub}
G_k=\{x\in \G^2_+: \ |x|/k\in\bn \}, \ \ k\geq 2
\end{equation}

One can check that $G_k$ is a sub-semigroup with a unit.

Let $\Phi=\{\z_1,\z_2,...,\z_q\}$, where $\z_1,\z_2,...,\z_q$ are
elements of ${\R}^{q-1}$ such that
\begin{equation}\label{21}
\z_i\z_j= \left\{ \begin{array}{ll}
1, \ \  \ \ \ \textrm{for $i=j$},\\
-\frac{1}{q-1}, \  \textrm{for $i\neq j$},
\end{array} \right.
\end{equation}
here $xy$, $x,y\in\br^{q-1}$, stands for the ordinary scalar
product on $\br^{q-1}$.

 From the last equality we infer that
\begin{equation}\label{zero}
 \sum_{k=1}^q\z_k=0.
\end{equation}

The vectors $\{\z_1,\z_2,...,\z_{q-1}\}$ are linearly independent,
therefore further they will be considered as a basis of
${\R}^{q-1}$.

In this paper we restrict ourselves to the case $q=3$. Then every
vector $h\in{\R}^2$ can be represented as $h=h_1\z_1+h_2\z_2$,
i.e. $h=(h_1,h_2)$, and from \eqref{21} we find
\begin{equation}\label{22}
h\z_i=\left\{
\begin{array}{lll}
h_1-\frac{1}{2}h_2, \qquad \ \ \ \ \textrm{if} \  \ i=1,\\
-\frac{1}{2}h_1+h_2, \qquad  \ \ \textrm{if} \  \ i=2,\\
 -\frac{1}{2}(h_1+h_2), \qquad \textrm{if} \ \ i=3.\\
\end{array}
\right.
\end{equation}

Let $\G^2_+=(V,\Lambda)$. We consider models where the spin takes
its values in the set $\Phi=\{\z_1,\z_2,\z_3\}$ and is assigned to
the vertices of the lattice $\G^2_+$. A configuration $\s$ on $V$
is then defined as a function $x\in V\to\s(x)\in\Phi$; in a
similar fashion one defines configurations $\s_n$ and $\s^{(n)}$
on $V_n$ and $W_n$, respectively. The set of all configurations on
$V$ (resp. $V_n$, $W_n$) coincides with $\Omega=\Phi^{V}$ (resp.
$\Omega_{V_n}=\Phi^{V_n},\ \ \Omega_{W_n}=\Phi^{W_n}$). One can
see that $\O_{V_n}=\O_{V_{n-1}}\times\O_{W_n}$. Using this, for
given configurations $\s_{n-1}\in\O_{V_{n-1}}$ and
$\s^{(n)}\in\O_{W_{n}}$ we define their concatenations  by the
formula

\begin{equation*}
\s_{n-1}\vee\s^{(n)}=\bigg\{\{\s_n(x),x\in
V_{n-1}\},\{\s^{(n)}(y),y\in W_n\}\bigg\}.
\end{equation*}

 It is clear that
$\s_{n-1}\vee\s^{(n)}\in \O_{V_n}$.

The Hamiltonian  of the Potts model with competing interactions
has the form

\begin{equation}\label{ham}
H(\s)=-J' \sum\limits_{>x,y<}\delta_{\s(x)\s(y)}-J_p\sum\limits_{
\widetilde{>x,y<}}\delta_{\s(x)\s(y)} -J_1'
\sum\limits_{<x,y>}\delta_{\s(x)\s(y)}\end{equation}
 where
$J',J_p,J_1'\in {\R}$ are coupling constants, $\s\in \Omega$ and
$\delta$ is the Kronecker symbol.

\section{The recurrent equations for the partition functions and Gibbs measures}

There are several approaches to derive an equation describing the
limiting Gibbs measures for the models on the Bethe lattices. One
approach is based on properties of Markov random fields, and
second one is based on recurrent equations for the partition
functions.

Recall that the total energy of a configuration $\s_n\in\O_{V_n}$
under condition $\bar\s_n\in\O_{V\setminus V_n}$ is defined by
$$
H(\s_n|\bar\s_n)=H(\s_n)+U(\s_n|\bar\s_n),
$$
here
\begin{eqnarray}
H(\s_n)&=&-J' \sum\limits_{\tiny{
\begin{array}{ll}
>x,y<\\
x,y\in V_n
\end{array}}
}\delta_{\s_n(x)\s_n(y)}-J_p\sum\limits_ {\tiny{
\begin{array}{ll}
\widetilde{>x,y<}\\
x,y\in V_n
\end{array}}}
\delta_{\s_n(x)\s_n(y)}\nonumber\\
&& -J_1' \sum\limits_{\tiny{
\begin{array}{ll}
<x,y>\\ x,y\in V_n
\end{array}}}\delta_{\s_n(x)\s_n(y)}\\\label{U}
U(\s_n|\bar\s_n)&=& -J' \sum\limits_{\tiny{
\begin{array}{lll}
>x,y<\\
x\in V_n,\\
y\in V\setminus V_n
\end{array}}
}\delta_{\s_n(x)\bar\s_n(y)}-J_p\sum\limits_ {\tiny{
\begin{array}{lll}
\widetilde{>x,y<}\nonumber\\
x\in V_n,\\
y\in V\setminus V_n
\end{array}}}
\delta_{\s_n(x)\bar\s_n(y)}\\
&&-J_1' \sum\limits_{\tiny{
\begin{array}{lll}
<x,y>\\
x\in V_n,\\
y\in V\setminus V_n
\end{array}}}\delta_{\s_n(x)\bar\s_n(y)}
\end{eqnarray}

The partition function $Z^{(n)}$ in volume $V_n$ under the
boundary condition $\bar\s_n$ is defined by
\begin{equation*}
Z^{(n)}=\sum_{\s\in\O_{V_n}}\exp(-\b H(\s|\bar\s_n)),
\end{equation*}
where $\b=1/T$ is the inverse temperature. Then the conditional
Gibbs measure $\m_{n}$ in volume $V_n$ under the boundary
condition $\bar\s_n$ is defined by
\begin{equation*}
\m_n(\s|\bar\s_n)=\frac{\exp(-\b H(\s|\bar\s_n))}{Z^{(n)}}, \ \
\s\in\O_{V_n}.
\end{equation*}

Consider $\O_{V_1}$  - the set of all configurations on
$V_1=\{(0),(1),(2)\}$, and enumerate all elements of it as shown
below:
\begin{eqnarray*}
\begin{array}{lll}
\sigma^{9(i-1)+1}=\{\z_i,\z_1,\z_1\},& \sigma^{9(i-1)+2}=\{\z_i,\z_1,\z_2\},&\sigma^{9(i-1)+3}=\{\z_i,\z_1,\z_3\}, \\
\sigma^{9(i-1)+4}=\{\z_i,\z_2,\z_1\},& \sigma^{9(i-1)+5}=\{\z_i,\z_2,\z_2\},&\sigma^{9(i-1)+6}=\{\z_i,\z_2,\z_3\}, \\
\sigma^{9(i-1)+7}=\{\z_i,\z_3,\z_1\},&
\sigma^{9(i-1)+8}=\{\z_i,\z_3,\z_2\},&\sigma^{9i}=\{\z_i,\z_3,\z_3\},
\end{array}
\end{eqnarray*}
where $i \in \{ 1,2,3 \}$.

We decompose the partition function $Z_{n}$ into 27 sums
$$
Z^{(n)}=\sum^{27}_{i=1}Z^{(n)}_{i},
$$
where
$$
Z^{(n)}_i=\sum\limits_{\sigma_n \in \Omega_{V_n} :
\sigma_n|_{V_1}=\sigma^i} {\exp(-\beta H_n(\sigma_n|\bar\s_n))},
\quad  i\in \{1,2,\cdots, 27\}.
$$

We set
$$
\theta=\exp(\beta J'); \quad \theta_p=\exp(\beta J_p); \quad
\theta_1=\exp(\beta J_1');
$$
and
\begin{eqnarray}\label{tZ}
\tilde Z^{(n)}_i=\sum\limits_{\sigma_n \in \Omega_{V_n} :
\sigma_n(0)=\z_i} {\exp(-\beta H_n(\sigma_n|\bar\s_n))}, \quad
i\in \{1,2,3\},
\end{eqnarray} that is
\begin{eqnarray*}
\tilde Z^{(n)}_i=\sum_{k=1}^9 Z^{(n)}_{9(i-1)+k}, \quad  i\in
\{1,2,3\}.
\end{eqnarray*}

Taking into account the denotation \eqref{ABC} through a direct
calculation one gets the following system of recurrent equations
\begin{eqnarray}\label{sys1}
\begin{array}{lll}
Z^{(n+1)}_{1}=\theta\theta^2_1 (A^{(n)}_{1})^2, & Z^{(n+1)}_{10}=\theta  (A^{(n)}_{2})^2, & Z^{(n+1)}_{19}=\theta(A^{(n)}_{3})^2 , \\
Z^{(n+1)}_{2}=\theta_1 A^{(n)}_{1}B^{(n)}_{1}, & Z^{(n+1)}_{11}=\theta_1 A^{(n)}_{2}B^{(n)}_{2}, & Z^{(n+1)}_{20}=A^{(n)}_{3}B^{(n)}_{3} , \\
Z^{(n+1)}_{3}=\theta_1 A^{(n)}_{1}C^{(n)}_{1}, & Z^{(n+1)}_{12}=A^{(n)}_{2}C^{(n)}_{2}, & Z^{(n+1)}_{21}=\theta_1 A^{(n)}_{3}B^{(n)}_{3} , \\
Z^{(n+1)}_{4}=Z^{(n+1)}_{2}, & Z^{(n+1)}_{13}=Z^{(n+1)}_{11}, & Z^{(n+1)}_{22}=Z^{(n+1)}_{20} , \\
Z^{(n+1)}_{5}=\theta (B^{(n)}_{1})^2, & Z^{(n+1)}_{14}=\theta \theta_1^2  (B^{(n)}_{2})^2, & Z^{(n+1)}_{23}=\theta(B^{(n)}_{3})^2 , \\
Z^{(n+1)}_{6}= B^{(n)}_{1}C^{(n)}_{1}, & Z^{(n+1)}_{15}= \theta_1  B^{(n)}_{2}C^{(n)}_{2}, & Z^{(n+1)}_{23}=\theta_1 B^{(n)}_{3} C^{(n)}_{3} , \\
Z^{(n+1)}_{7}=Z^{(n+1)}_{3}, & Z^{(n+1)}_{16}=Z^{(n+1)}_{12}, & Z^{(n+1)}_{25}=Z^{(n+1)}_{21} , \\
Z^{(n+1)}_{8}=Z^{(n+1)}_{6}, & Z^{(n+1)}_{17}=Z^{(n+1)}_{15}, & Z^{(n+1)}_{26}=Z^{(n+1)}_{24} , \\
Z^{(n+1)}_{9}=\theta  (C^{(n)}_{1})^2, & Z^{(n+1)}_{18}=\theta  (C^{(n)}_{2})^2, & Z^{(n+1)}_{27}=\theta \theta^2_1 (C^{(n)}_{3})^2 . \\
\end{array}
\end{eqnarray}

Introducing new variables
\begin{eqnarray}\label{den}
\begin{array}{lll}
x^{(n)}_1=Z^{(n)}_{1}; &x^{(n)}_2=Z^{(n)}_{2}=Z^{(n)}_{4}; & x^{(n)}_3=Z^{(n)}_{3}=Z^{(n)}_{7};\\
x^{(n)}_4=Z^{(n)}_{5} ;&x^{(n)}_5=Z^{(n)}_{6}=Z^{(n)}_{8}; & x^{(n)}_6=Z^{(n)}_{9} \\
x^{(n)}_7=Z^{(n)}_{10}; &x^{(n)}_8=Z^{(n)}_{11}=Z^{(n)}_{13}; & x^{(n)}_9=Z^{(n)}_{12}=Z^{(n)}_{16}; \\
x^{(n)}_{10}=Z^{(n)}_{14} ;&x^{(n)}_{11}=Z^{(n)}_{15}=Z^{(n)}_{17}; & x^{(n)}_{12}=Z^{(n)}_{18} \\
x^{(n)}_{13}=Z^{(n)}_{19};
&x^{(n)}_{14}=Z^{(n)}_{20}=Z^{(n)}_{22}; &
x^{(n)}_{15}=Z^{(n)}_{21}=Z^{(n)}_{25};
\\
x^{(n)}_{16}=Z^{(n)}_{23}
;&x^{(n)}_{17}=Z^{(n)}_{24}=Z^{(n)}_{26}; &
x^{(n)}_{18}=Z^{(n)}_{27}
\end{array}
\end{eqnarray}
the equations \eqref{sys1} are represented by
\begin{eqnarray}\label{rec11}
\left\{
\begin{array}{ll}
x^{(n+1)}_1= \theta\theta^2_1 (A^{(n)}_1)^2 &
x^{(n+1)}_2= \theta_1 A^{(n)}_1B^{(n)}_1, \\
x^{(n+1)}_3= \theta_1 A^{(n)}_1C^{(n)}_1, &
x^{(n+1)}_4= \theta (B^{(n)}_1)^2,\\
x^{(n+1)}_5 = B^{(n)}_1 C^{(n)}_1, &
x^{(n+1)}_6 = \theta (C^{(n)}_1)^2, \\
x^{(n+1)}_7 = \theta (A^{(n)}_2)^2, &
x^{(n+1)}_8 =  \theta_1A^{(n)}_2 B^{(n)}_2,  \\
x^{(n+1)}_9 =  A^{(n)}_2 C^{(n)}_2, &
x^{(n+1)}_{10} = \theta\theta_1^2  (B^{(n)}_2)^2,  \\
x^{(n+1)}_{11} = \theta_1  B^{(n)}_2 C^{(n)}_2 , &
x^{(n+1)}_{12} = \theta (C_2^{(n)})^2 \\
x^{(n+1)}_{13} = \theta (A_3^{(n)})^2,  &
x^{(n+1)}_{14} = A^{(n)}_3 B^{(n)}_3,   \\
x^{(n+1)}_{15} = \theta_1 A_3^{(n)} C_3^{(n)}, &
x^{(n+1)}_{16} = \theta (B_3^{(n)})^2, \\
x^{(n+1)}_{17} = \theta_1  B^{(n)}_3 C^{(n)}_3,  & x^{(n+1)}_{18}
= \theta\theta_1^2  (C_3^{(n)})^2.\\
\end{array}
\right.
\end{eqnarray}
The asymptotic behavior of the recurrence system \eqref{rec11} is
defined by the first date $\{x^{(1)}_k:\ k=1,2,\dots, 18\}$, which
is  in turn determined by a boundary condition $\bar\s$.

Let us separately consider free boundary condition, that is
$U(\s|\bar\s)$ is zero, and three boundary conditions
$\bar\s_n\equiv\z_i$, where $i=1,2,3$. Here by $\bar\s_n\equiv\z$
we have meant a configuration defined by $\bar\s_n=\{\s(x):
\s(x)=\z, \forall x\in V\setminus V_n\}$.

For the free boundary we have
\begin{eqnarray*}
\begin{array}{lll}
x^{(1)}_1=\theta\theta_1^2; &x^{(1)}_2=\theta_1; & x^{(1)}_3=\theta_1;\\
x^{(1)}_4=\theta ;& x^{(1)}_5=1; & x^{(1)}_6=\theta; \\
x^{(1)}_7=\theta; &x^{(1)}_8=\theta_1; & x^{(1)}_9=1; \\
x^{(1)}_{10}=\theta\theta_1^2 ;&x^{(1)}_{11}=\theta_1; & x^{(1)}_{12}=\theta; \\
x^{(1)}_{13}=\theta; &x^{(1)}_{14}=1; & x^{(1)}_{15}=\theta_1; \\
x^{(1)}_{16}=\theta ;&x^{(1)}_{17}=\theta_1; &
x^{(1)}_{18}=\theta\theta_1^2
\end{array}
\end{eqnarray*}
and from the direct calculations (see \eqref{ABC1}) we infer that
\begin{eqnarray*}
&&A_1^{(n)}=B_2^{(n)}=C_3^{(n)},\\
&&A_2^{(n)}=A_3^{(n)}=B_1^{(n)}=B_3^{(n)}=C_1^{(n)}=C_2^{(n)},
\end{eqnarray*}
so that
$$
\tilde Z_1^{(n)}=\tilde Z_2^{(n)}=\tilde Z_3^{(n)}.
$$
Hence the corresponding Gibbs measure $\m_0$ is the {\it unordered
phase}, i.e. $\m(\s(x)=\z_i)=1/3$ for any $x\in \G^2_+$,
$i=1,2,3$.

Now consider boundary condition $\bar\s\equiv\z_1$. Then we have
\begin{eqnarray*}
\begin{array}{lll}
x^{(1)}_1=\theta\theta_1^6\theta_p^4; &x^{(1)}_2=\theta_1^3\theta_p^4; & x^{(1)}_3=\theta_1^3\theta_p^4;\\
x^{(1)}_4=\theta\theta_p^4 ;& x^{(1)}_5=\theta_p^4; & x^{(1)}_6=\theta\theta_p^4; \\
x^{(1)}_7=\theta\theta_1^4; &x^{(1)}_8=\theta_1^3; & x^{(1)}_9=\theta_1^2; \\
x^{(1)}_{10}=\theta\theta_1^2 ;&x^{(1)}_{11}=\theta_1; & x^{(1)}_{12}=\theta; \\
x^{(1)}_{13}=\theta\theta_1^4; &x^{(1)}_{14}=\theta_1^2; & x^{(1)}_{15}=\theta_1^3; \\
x^{(1)}_{16}=\theta ;&x^{(1)}_{17}=\theta_1; &
x^{(1)}_{18}=\theta\theta_1^2.
\end{array}
\end{eqnarray*}
By simple calculations (see \eqref{ABC1}) we obtain
\begin{eqnarray*}
&&B_1^{(n)}=C_1^{(n)}, \quad A_2^{(n)}=A_3^{(n)},\\
&&B_2^{(n)}=C_3^{(n)}, \quad B_3^{(n)}=C_2^{(n)},
\end{eqnarray*}
and
\begin{eqnarray*}
\tilde Z_1^{(n+1)}&=& \theta\theta_1^2(A_1^{(n)})^2+4\theta_1A_1^{(n)}B_1^{(n)}+2(\theta+1)(B_1^{(n)})^2,\\
\tilde Z_2^{(n+1)}= \tilde
Z_3^{(n+1)}&=&\theta(A_2^{(n)})^2+2\theta_1A_2^{(n)}B_2^{(n)}+2A_2^{(n)}C_2^{(n)}\\
&&+\theta\theta_1^2(B_2^{(n)})^2+
2\theta_1B_2^{(n)}C_2^{(n)}+\theta(C_2^{(n)})^2.
\end{eqnarray*}

By the same argument for the boundary condition $\bar\s\equiv\z_2$
we have
$$
\tilde Z_1^{(n)}= \tilde Z_3^{(n)}
$$
and for the boundary condition  $\bar\s\equiv\z_3$
$$
\tilde Z_1^{(n)}= \tilde Z_2^{(n)}.
$$

If $\theta_p=1$, i.e. $J_p=0$, then from the system of equations
\eqref{sys1} we derive
\begin{eqnarray}\label{zz}
\begin{array}{lll}
\tilde Z^{(n+1)}_{1} = \theta\theta_1^2  (\tilde Z^{(n)}_{1})^2 +
2 \theta_1  \tilde Z^{(n)}_{1} \tilde Z^{(n)}_{2}+ 2 \theta_1
\tilde Z^{(n)}_{1} \tilde Z^{(n)}_{3} + \theta (\tilde
Z^{(n)}_{2})^2 +
2 \theta_1  \tilde Z^{(n)}_{2} \tilde Z^{(n)}_{3}+ \theta(\tilde Z^{(n)}_{3})^2  \\
\tilde Z^{(n+1)}_{2} = \theta (\tilde Z^{(n)}_{1})^2  + 2 \theta_1
\tilde Z^{(n)}_{1} \tilde Z^{(n)}_{2}+ 2  \tilde Z^{(n)}_{1}
\tilde Z^{(n)}_{3} + \theta\theta_1^2 (\tilde Z^{(n)}_{2})^2 +
2 \theta_1  \tilde Z^{(n)}_{2} \tilde Z^{(n)}_{3}+ \theta(\tilde Z^{(n)}_{3})^2  \\
\tilde Z^{(n+1)}_{3} = \theta(\tilde Z^{(n)}_{1})^2  + 2 \tilde
Z^{(n)}_{1} \tilde Z^{(n)}_{2}+ 2  \theta_1 \tilde Z^{(n)}_{1}
\tilde Z^{(n)}_{3} + \theta (\tilde Z^{(n)}_{2})^2 + 2 \theta_1
\tilde Z^{(n)}_{2} \tilde Z^{(n)}_{3}+ \theta\theta_1^2
(\tilde Z^{(n)}_{3})^2
\end{array}
\end{eqnarray}

Letting
\begin{equation*}
u_n=\frac{\tilde Z^{(n)}_{1} }{ \tilde Z^{(n)}_{3}} \quad
\mbox{and} \quad v_n=\frac{\tilde Z^{(n)}_{2} }{ \tilde
Z^{(n)}_{3}}.
\end{equation*}
then from \eqref{zz} one gets
\begin{eqnarray}\label{rec0}
\left\{
\begin{array}{ll}
u_{n+1}=\frac{\theta \theta_1^2  u_n^2 +2\theta_1 u_nv_n +  \theta
v_n^2+2 \theta_1 u_n + 2 v_n + \theta}{\theta u_n^2 +2 u_nv_n
+\theta v_n^2+ 2 \theta_1 u_n +  2\theta_1 v_n + \theta
\theta_1^2},
\\[3mm]
v_{n+1}=\frac{\theta u_n^2 +2\theta_1 u_nv_n +  \theta \theta_1^2
v_n^2 +  2 u_n +2 \theta_1 v_n + \theta}{\theta u_n^2 +2 u_nv_n +
 \theta v_n^2 +2 \theta_1 u_n + 2\theta_1 v_n + \theta
\theta_1^2}.
\end{array}
\right.
\end{eqnarray}

From the above made statements we conclude that
\begin{itemize}
    \item[(i)] $u_n=v_n=1$, $\forall n\in\bn$ for the free boundary
condition;
    \item[(ii)] $v_n=1$, $\forall n\in\bn$ for the boundary
condition $\bar\s\equiv\z_1$;
    \item[(iii)] $u_n=1$, $\forall n\in\bn$ for the boundary condition $\bar\s\equiv\z_2$;
    \item[(iv)] $u_n=v_n$, $\forall n\in\bn$ for the boundary condition
$\bar\s\equiv\z_3$ .
\end{itemize}

Consequently, when $J_p=0$ we can receive an exact solution. In
the next section we will find an exact critical curve and the free
energy for this case.\\

Now let us assume that $J_p\neq 0$ and $\bar\s\equiv\z_1$. Then
the system \eqref{rec11} reduces to a system consisting of five
independent variables (see Appendix A), but a new recurrence
system still remains rather complicated . Therefore, it is natural
to begin our investigation with the case $J_p=0$. In the case
$J_p\neq 0$ a full analysis of such a system will be a theme of
our next investigations \cite{GMMP}, where the modulated phases
and Lifshitz points will be discussed.\\

Now we are going to show how the equations \eqref{rec0} are
related with the surface energy \eqref{U} of the given
Hamiltonian. To do it, we give a construction of a special class
of limiting Gibbs measures for the model when $J_p=0$.

Let us note that the equality \eqref{21} implies that
\begin{equation*}
\delta_{\sigma(x)\sigma(y)}=\frac{2}{3}
\bigg(\sigma(x)\sigma(y)+\frac{1}{2}\bigg)
\end{equation*}
for all $ x,y \in V $. Therefore, the Hamiltonian $H(\sigma)$ is
rewritten by
\begin{equation}\label{ham1}
H(\sigma)=-J\sum_{>x,y<}\sigma(x)\sigma(y)-J_1
\sum\limits_{<x,y>}\s(x)\s(y), \end{equation} where
$J=\dsf{2}{3}J', J_1=\dsf{2}{3}J_1'$.

Let ${\mathbf{h}}:x\to h_x=(h_{1,x},h_{2,x})\in{\R}^{2}$ be a real
vector-valued function of $x\in V$. Given $n=1,2,...$ consider the
probability measure $\m^{(n)}$ on $\Phi^{V_n}$ defined by
\begin{equation}\label{mes1}
\mu^{(n)}(\s_n)=(Z^{(n)})^{-1}\exp\{-\b H(\s_n)+\sum_{x\in W_n}
h_x\s_n(x)\},
\end{equation} where
\begin{equation*}
 H(\s_n)=-J \sum\limits_{>x,y<: x,y\in V_n}{\s_n(x)\s_n(y)} -J_1
\sum\limits_{<x,y>: x,y\in V_n}{\s_n(x)\s_n(y)}, \end{equation*}
and as before $\b=\frac{1}{T}$ and  $\s_n\in \O_{V_n}$ and
$Z^{(n)}$ is the corresponding partition function:
\begin{equation}\label{partition}
Z^{(n)}\equiv Z^{(n)}(\b,h)=\sum_{\tilde\s_n\in\Omega_{V_n}}\exp
\{-\b H(\tilde\s_n)+\sum_{x\in W_n}h_x\tilde\s_n(x)\}.
\end{equation}

Let $V_1\subset V_2\subset...$
$\bigcup\limits_{n=1}^{\infty}V_n=V$ and $\m^{(1)},\m^{(2)},...$
be a sequence of probability measures on
$\Phi^{V_1},\Phi^{V_2},...$ given by  \eqref{mes1}. If these
measures satisfy the consistency condition
\begin{equation}\label{consis1}
\sum_{\s^{(n)}}\m^{(n)}(\s_{n-1}\vee\s^{(n)})=\m^{(n-1)}(\s_{n-1}),
\end{equation}
where $\s^{(n)}=\{\s(x), x\in W_n\}$, then according to the
Kolmogorov theorem, (see, e.g. Ref. \cite{Sh}) there is a unique
limiting Gibbs measure $\m$ on $(\O,\mathcal{F})$, where
$\mathcal{F}$ is a $\s$-algebra generated by cylindrical subset of
$\O$, such that for every $n=1,2,...$ and $\s_n\in\Phi^{V_n}$ the
following equality holds
\begin{equation*}
\m\bigg(\{\s|_{V_n}=\s_n\}\bigg)=\m^{(n)}(\s_n).
\end{equation*}

One can see that the consistency condition \eqref{consis1} is
satisfied if and only if the function ${\mathbf{h}}$ satisfies the
following equation
\begin{equation}\label{35}
\left\{
\begin{array}{ll}
h_{x,1}'= \log F(h_y',h_z')\\
h_{x,2}'= \log F((h_y')^t,(h_z')^t), \end{array} \right.
\end{equation}
here and below for given vector $h=(h_1,h_2)$ by $h'$ and $h^t$ we
have denoted the vectors $\frac{3}{2}h$ and $(h_2,h_1)$
respectively, and $F:{\R}^{q-1}\times {\R}^{q-1}\to{\R}$ is a
function defined by
\begin{equation}\label{func1}
F(h,r)=\frac{\theta_1^2\theta e^{h_1+r_1}+\theta_1(e^{h_1+r_2}+
e^{h_2+r_1})+\theta
e^{h_2+r_2}+\theta_1(e^{h_1}+e^{r_1})+e^{h_2}+e^{r_2}+\theta}
{\theta e^{h_1+r_1}+e^{h_1+r_2}+ e^{h_2+r_1}+\theta
e^{h_2+r_2}+\theta_1(e^{h_1}+e^{r_1}+e^{h_2}+e^{r_2})+\theta_1^2\theta}
\end{equation}
where $h=(h_1,h_2), r=(r_1,r_2)$ and $<y,x,z>$ are ternary
neighbors (see Appendix B for the proof).

Consequently, the problem of describing the Gibbs measures is
reduced to the description of solutions of the functional equation
\eqref{35}. On the other hand, we see that from the derived
equation \eqref{35} we can obtain  \eqref{rec0}, when the function
${\mathbf{h}}$ is translation invariant.

\section{Ground states of the model}

In this section we are going to describe ground states of the
model. Recall that a relative Hamiltonian $H(\s,\f)$ is defined by
the difference between the energies of configurations $\s, \f$
\begin{equation}\label{reham1}
H(\s,\f)=-J'\sum_{>x,y<}(\delta_{\sigma(x)\sigma(y)}-\delta_{\f(x)\f(y)})-
J_1'\sum_{<x,y>}(\delta_{\sigma(x)\sigma(y)}-\delta_{\f(x)\f(y)}),
\end{equation}
where $J=(J',J_1')\in{\R}^2$ is an arbitrary fixed parameter.

In the sequel as usual we denote the cardinality number of a set
$A$ by $|A|$. A set $c$ consisting of three vertices
$\{x_1,\{x_2,x_3\}\}$ is called a {\it cell} if these vertices are
$<x_2,x_1,x_3>$ ternary. In this case, the vertex $x_1$ is called
{\it the origin} of a cell $c$. By $\mathcal{C}$ the set of all
cells is denoted. We say that two $c$ and $c'$ cells are {\it
nearest neighbor} if $|c\cap c'|=1$, and denote them by $<c,c'>$.
From this definition we see that if $c$ and $c'$ cells are not
nearest neighbor then either they coincide or disjoint.  Let
$\s\in\O$ and $c\in\cc$, then the restriction of a configuration
$\s$ to $c$ is denoted by $\s_c$, and we will use to write
elements of $\s_c$ as follows
\begin{equation*}
\s_c=\{\s(x_1),\{\s(x_2),\s(x_3)\}\}.
\end{equation*}
The set of all configurations on $c$ is denoted by $\O_c$.

The energy of a cell $c$ at a configuration $\s$ is defined by
\begin{equation}\label{ham11}
U(\s_c)=-J'\sum\limits_{>x,y< : x,y\in c} \d_{\s(x)\s(y)}-J_1'
\sum\limits_{<x,y>: x,y\in c}\d_{\s(x)\s(y)}.
\end{equation}

From \eqref{ham11} one can deduce that for any $c\in\cc$ and
$\s\in\O$ we have
\[
U(\s_c)\in \{U_1(J), U_2(J),U_3(J), U_{4}(J)\}, \] where
\begin{equation}\label{U}
 U_1(J)=-2J_1'-J',
 \ \ U_2(J)=-J'_1, \ \ U_3(J)=-J', \ \ U_4(J)=0, \ \ J=(J',J_1').
\end{equation}

Denote
$$
{\cb}_i=\{\s_c\in \O_c : U(\s_c)=U_i\}, \ \ \  i=1,2,3,4,
$$
then using a combinatorial calculation one can show the following

\begin{eqnarray}\label{gr}
{\cb}_1&=&\bigg\{\{\z_i,\{\z_i,\z_i\}\}, \ i=1,2,3 \bigg\},\\
\label{1gr} {\cb}_2&=&\bigg\{\{\z_i,\{\z_i,\z_j\}\},
\{\z_i,\{\z_j,\z_i\}\}, \ i\neq j, \ i,j\in\{1,2,3\} \bigg\},\\
\label{2gr}{\cb}_3&=&\bigg\{\{\z_j,\{\z_i,\z_i\}\}, \ i\neq j, \
i,j\in\{1,2,3\} \bigg\},\\ \label{3gr}
{\cb}_4&=&\bigg\{\{\z_i,\{\z_j,\z_k\}\}, \ i,j,k\in\{1,2,3\}, \
i\cdot j\cdot k=6 \bigg\}. \end{eqnarray}

From \eqref{reham1} we infer that
\begin{equation}\label{reham2}
H(\f, \s)=\sum_{c\in\cc}(U(\f_c)-U(\s_c)).
\end{equation}

Recall (see \cite{R}) that a  configuration $\f\in\O$ is called a
{\it ground state} for the relative Hamiltonian of $H$ if
\begin{equation}\label{gs}
U(\f_c)=\min\{U_1(J), U_2(J),U_3(J), U_{4}(J)\},\ \  \mbox{ for
any} \ \ c\in\cc.
\end{equation}

A couple of configurations $\s,\f\in\Omega$ coincide {\it almost
everywhere}, if they are different except for a finite number of
positions and which are denoted by $\s=\f$ [a.s].

\begin{prop}\label{gs1} A configuration $\f$ is a ground state for $H$ if and only if
the following inequality holds
\begin{equation}\label{f-s}
H(\f,\s)\leq 0
\end{equation}
for every $\s\in\O$ with  $\s=\f$ [a.s].
\end{prop}

\begin{proof} The almost every coincidence of $\s$ and $\f$
implies that there exists a finite subset $L\subset V$ such that
$\s(x)\neq\f(x)$ for all $x\in L$. Denote
$V_L=\bigcap\limits_{k=1}^{\infty}\{V_k: L\subset V_k\}$. Then
taking into account that $\f$ is a ground state we have
$U(\f_c)\leq U(\s_c)$ for every $c\in\cc$. So, using the last
inequality and \eqref{reham2} one gets
\begin{equation*}
H(\f, \s)=\sum_{c\in\cc,c\in V_L}(U(\f_c)-U(\s_c))\leq 0.
\end{equation*}

Now assume that \eqref{f-s} holds. Take any cell $c\in \cc$.
Consider the following configuration:
$$
\s_{c,\f}(x)= \left\{
\begin{array}{ll}
\s(x), \ \ \ \textrm{if} \ \ x\in c,\\
\f(x), \ \ \ \textrm{if} \ \ x\notin c,
\end{array}
\right.
$$
where $\s\in\O_c$. It is clear that $\s_{c,\f}=\f$ [a.s.], so from
\eqref{reham2} and \eqref{f-s} we infer that
$H(\f,\s_{c,\f})=U(\f_c)-U(\s)\leq 0$, i.e. $U(\f_c)\leq U(\s)$.
From the arbitrariness of $\s$ one finds that $\f$ is a ground
state. \end{proof} Denote
\begin{equation*}
A_k=\bigg\{J\in\br^2: U_k(J)=\min\{U_1(J), U_2(J),U_3(J),
U_{4}(J)\}\bigg\}, \ \ k=1,2,3,4.
\end{equation*}

From equalities \eqref{U} we can easily get the following
\begin{eqnarray*}
A_1&=&\{J=(J',J_1')\in\br^2: J_1'\geq 0,\ J_1'+J'\geq 0\}\\
A_2&=&\{J=(J',J_1')\in\br^2: J_1'\geq 0,\
J_1'+J'\leq 0\}\\
A_3&=&\{J=(J',J_1')\in\br^2: J_1'\leq 0,\ J'>0 \}\\
A_4&=&\{J=(J',J_1')\in\br^2: J_1'\leq 0,\ J'<0\}
\end{eqnarray*}

Denote
\begin{equation*}
B_k=A_k\setminus\bigg(\bigcup_{j=1}^4A_k\cap A_j\bigg), \ \
k=1,2,3,4.
\end{equation*}

Now we are are going to construct the ground states for the model.
Before doing it let us introduce some notions. Take two nearest
neighbor cells $c,c'\in\cc$ with common vertex $x\in c\cap c'$. We
say that two configurations $\s_c\in\O_c$ and $\s_{c'}\in\O_{c'}$
are {\it consistent} if $\s_c(x)=\s_{c'}(x)$. It is easy to see
that the set $V$ can be represented as a union of all nearest
neighbor cells, therefore to define a configuration $\s$ on whole
$V$, it is enough to determine one on nearest neighbor cells such
that its values should be consistent on such cells. Namely, each
configuration $\s\in\O$ is represented as a family of consistent
configurations on $\O_c$, i.e. $\s=\{\s_c\}_{c\in\cc}$. Therefore,
from the definition of the ground state and \eqref{gr}-\eqref{3gr}
we are able to formulate the following

\begin{prop}\label{1gs} Let $J\in B_k$ then a configuration $\f=\{\f_c\}_{c\in\cc}$ is a ground state
if and only if $\f_c\in\cb_k$ for all $c\in\cc$.
\end{prop}

Let us denote
\begin{equation}\label{gs2}
\s^{(m)}=\{\s(x): \s(x)=\z_m, \ \forall x\in V\}, \ \ m=1,2,3.
\end{equation}

\begin{thm}\label{gs3} Let $J\in B_i$, then for any fixed
$\s_c\in \cb_i$ (here $c$ is fixed), there exists a ground state
$\f\in\O$ with $\f_c=\s_c.$
\end{thm}
\begin{proof} Let $\s_c\in \cb_i$. Without loss of generality we may assume that the
center $x_1$ of $c$ is the origin of the lattice $\G^2_+$. Further
we will suppose that $\s(x_1)=\z_1$ (other cases are similarly
proceeded). Put
\begin{eqnarray*}
N_{j}^{(i)}(\s_c)&=&\left|\bigg\{k\in\{1,2,3\}:
\s_c(x_k)=\z_j\bigg\}\right| \ \
j=1,2,3,\\
\bar
n_{i}(\s_c)&=&\bigg(N_{1}^{(i)}(\s_c),N_{2}^{(i)}(\s_c),N_{3}^{(i)}(\s_c)\bigg),
\ \ c\in\cc.
\end{eqnarray*}

It is clear that $N_{j}^{(i)}(\s_c)\geq 0$ and
$\sum\limits_{k=1}^3N_{k}^{(i)}(\s_c)=3$.

According to Proposition \ref{1gs} to find a ground state
$\f\in\O$ it is enough to construct a consistent family of ground
states $\{\f_c\}_{c\in\cc}$.

Consider several cases with respect to $i$ ($i\in\{1,2,3,4\}$).

{\tt Case $i=1$.} In this case, according to \eqref{gr} we have
$\s_c(x)=\z_1$ for every $x\in c$. This means that $\bar
n_{1}(\s_c)=(3,0,0)$. Then the configuration $\s^{(1)}$ is the
required one and it is a ground state. From \eqref{trans1} we see
that $\s^{(1)}$ is translation-invariant.

{\tt Case $i=2$.} In this case from \eqref{1gr} we find that $\bar
n_{2}(\s_c)$ is either $(2,0,1)$ or $(2,1,0)$. Let us assume that
$\bar n_{2}(\s_c)=(2,0,1)$. Now we want to construct a ground
state on nearest neighbor cells, therefore take $c',c''\in\cc$
such that $<c,c'>$, $<c,c''>$ and $c'\neq c''$. It is clear that
$c'\cap c''=\emptyset$. Let  $x_2$ and $x_3$ be the centers of
$c'$ and $c''$, respectively. So due to our assumption we find
that either $\s(x_2)=\z_1$, $\s(x_3)=\z_3$ or $\s(x_2)=\z_3$,
$\s(x_2)=\z_1$. Let us consider $\s(x_2)=\z_1$, $\s(x_3)=\z_3$.
Then we have $\s_{c}=\{\z_1,\{\z_1,\z_3\}\}$. We are going to
determine configurations $\f_{c'}\in\O_{c'}$,
$\f_{c''}\in\O_{c''}$ consistent with $\s_c$ and
$N_{1}^{(2)}(\s)\cdot N_{3}^{(2)}(\s)=2$, $\s=\f{_c'},\f{_c''}$.
To do it, by means of \eqref{1gr}, we choose configurations $\f_c$
and $\f_{c'}$ on $c', c''$, respectively, as follows
\begin{equation}\label{conf1}
\f_{c'}=\{\z_1,\{\z_1,\z_3\}\}, \ \
\f_{c''}=\{\z_3,\{\z_1,\z_3\}\}.
\end{equation}
Hence continuing this procedure one can construct a configuration
$\f$ on $V$, and denote it by $\f^{(1,3)}$. From the construction
we infer that $\f^{(1,3)}$ satisfies the required conditions (see
Fig. \ref{fig2}). The constructed configuration is quasi
$\G^2_+$-periodic. Indeed, from \eqref{trans2} and \eqref{conf1}
one can check that for every $x\in \G^2_+$ with $|x|\neq 1$ we
have $\f^{(1,3)}(\pi^{(\g)}_{(0)}(x))=\f^{(1,3)}(x)$, here
$\g(\{1,2\})=\{2,1\}$. So from \eqref{trans3} for every $g\in
\G^2_+$ one finds that
$\f^{(1,3)}(\pi^{(\g)}_{g}(x))=\f^{(1,3)}(x)$ for all $|x|\neq 1$.
Similarly, we can construct the following quasi periodic ground
states:
$$
\f^{(3,1)},\f^{(1,2)},\f^{(2,1)},\f^{(2,3)},\f^{(3,2)}.
$$

\begin{figure}
\begin{center}
\includegraphics[width=10.07cm]{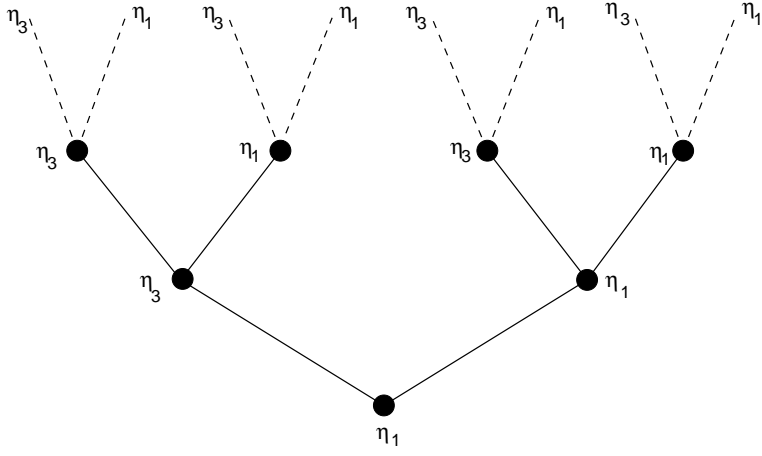}
\end{center}
\caption{$\f^{(1,3)}-$ ground state. The coupling constants belong
to $B_2$} \label{fig2}
\end{figure}

{\tt Case $i=3$.} In this setting we have that $\bar n_{3}(\s_c)$
is either $(1,0,2)$ or $(1,2,0)$ (see \eqref{2gr}). Let us assume
that $\bar n_{2}(\s_c)=(1,2,0)$. Let $c',c''\in\cc$ be as above.
From  \eqref{2gr} and our assumption one finds
$\s(x_2)=\s(x_3)=\z_2$. Then again taking into account \eqref{2gr}
for $c', c''$ we can define consistent configurations by
\begin{equation}\label{conf2}
\f_{c'}=\{\z_2,\{\z_1,\z_1\}\}, \ \
\f_{c''}=\{\z_2,\{\z_1,\z_1\}\}.
\end{equation}
Again continuing this procedure we obtain a configuration on $V$,
which we denote by $\f^{[1,2]}$. From the construction we infer
that $\f^{[1,2]}$ is a ground state and satisfies the needed
conditions (see Fig.\ref{fig3}). From \eqref{conf2} and
\eqref{trans1} we immediately conclude that it is $G_2$-periodic.
Similarly, we can construct the following $G_2$-periodic ground
states:
$$
\f^{[2,1]},\f^{[1,3]},\f^{[3,1]},\f^{[2,3]},\f^{[3,2]}.
$$

\begin{figure}
\begin{center}
\includegraphics[width=10.07cm]{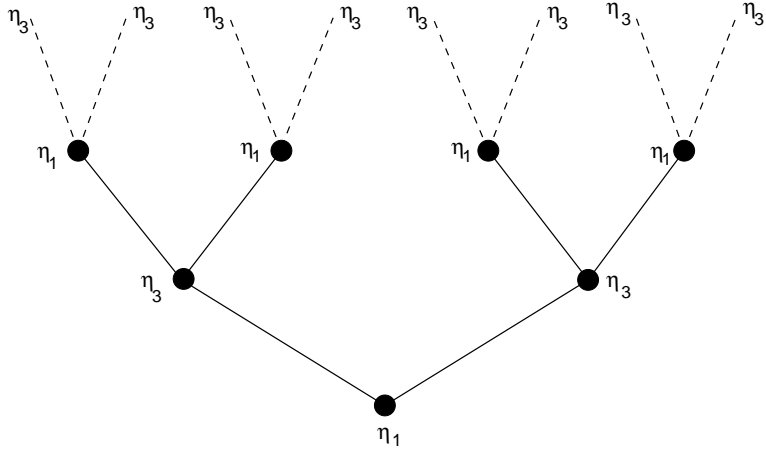}
\end{center}
\caption{$\f^{[1,3]}-$ ground state. The coupling constants belong
to $B_3$} \label{fig3}
\end{figure}
Note that on $c', c''$ we also may determine another consistent
configurations by
\begin{equation}\label{conf3}
\f_{c'}=\{\z_2,\{\z_3,\z_3\}\}, \ \
\f_{c''}=\{\z_2,\{\z_3,\z_3\}\}.
\end{equation}
Now take $b',b''\in\cc$ such that $<c',b'>$, $<c',b''>$ and
$b'\neq b''$. On $b', b''$ we define consistent configurations
with $\f_{c'}$ by
\begin{equation}\label{conf4}
\f_{b'}=\{\z_3,\{\z_1,\z_1\}\}, \ \
\f_{b''}=\{\z_3,\{\z_1,\z_1\}\}.
\end{equation}
Analogously, one defines $\f$ on the neighboring cells of $c''$.
Consequently, continuing this procedure we construct a
configuration $\f^{[1,2,3]}$ on $V$. From
\eqref{sub},\eqref{trans1},\eqref{conf3} and \eqref{conf4} we see
that $\f^{[1,2,3]}$ is a $G_3$-periodic ground state. Similarly,
reasoning one can be built the following $G_3$-periodic ground
states:
$$
\f^{[2,1,3]},\f^{[2,3,1]},\f^{[1,3,2]},\f^{[3,1,2]},\f^{[3,2,1]}.
$$

These constructions lead us to make a conclusion that for any
number of collection $\{i_1,\dots,i_k\}$ with $i_m\neq i_{m+1}$,
$i_m\in\{1,2,3\}$ we may construct a ground state
$\f^{[i_1,\dots,i_k]}$ which is $G_k$-invariant. Hence, there are
countable number periodic ground states.

{\tt Case $i=4$}. In this case using the same argument as in the
previous cases we can construct a required ground state, but it
would be non-periodic (see \eqref{3gr}).
\end{proof}

{\bf Remark 1.} From the proof of Theorem \ref{gs3} one can see
that for a given $\s_c\in \cb_i$ with $i\geq 2$, there exist
continuum number of ground states $\f\in\O$ such that $\f_{c'}\in
\cb_i$ for any $c'\in\cc$ and $\f_c=\s_c.$ Since, in those cases
at each step we had two possibilities there have been  at least
two possibilities to choice of $\f_{c'}$ and $\f_{c''}$, this
means that a configuration on $V$ can be constructed by the
continuum number of ways.

\begin{cor}\label{gs4}  Let $J\in B_i$($i\neq 4$), then for any
fixed $\s_c\in \cb_i$ (here $c$ is fixed), there exists a periodic
(quasi) ground state $\f\in\O$ such that $\f_c=\s_c.$
\end{cor}

By $GS(H)$ and $GS_p(H)$ we denote the set of all ground states
and periodic ground states of the model \eqref{ham}, respectively.
Here by periodic configuration we mean $G$-periodic or quasi
$G$-periodic ones.

\begin{cor}\label{gs5}  For the Potts model \eqref{ham} the following assertions hold.
\begin{enumerate}
    \item[(i)] Let $J\in B_1$, then
\begin{equation*}
|GS(H)|=|GS_p(H)|=3;
\end{equation*}
    \item[(ii)] Let $J\in B_2$ then
\begin{equation*}
|GS(H)|=c, \ \ |GS_p(H)|=6;
\end{equation*}
  \item[(iii)] Let $J\in B_3$ then
\begin{equation*}
|GS(H)|=c, \ \ |GS_p(H)|=\aleph_0;
\end{equation*}
  \item[(iv)] Let $J\in B_4$ then
\begin{equation*}
|GS(H)|=c.
\end{equation*}
\end{enumerate}
\end{cor}

The proof immediately follows from Theorem \ref{gs3} and Remark 1.

\begin{figure}
\begin{center}
\includegraphics[width=10.07cm]{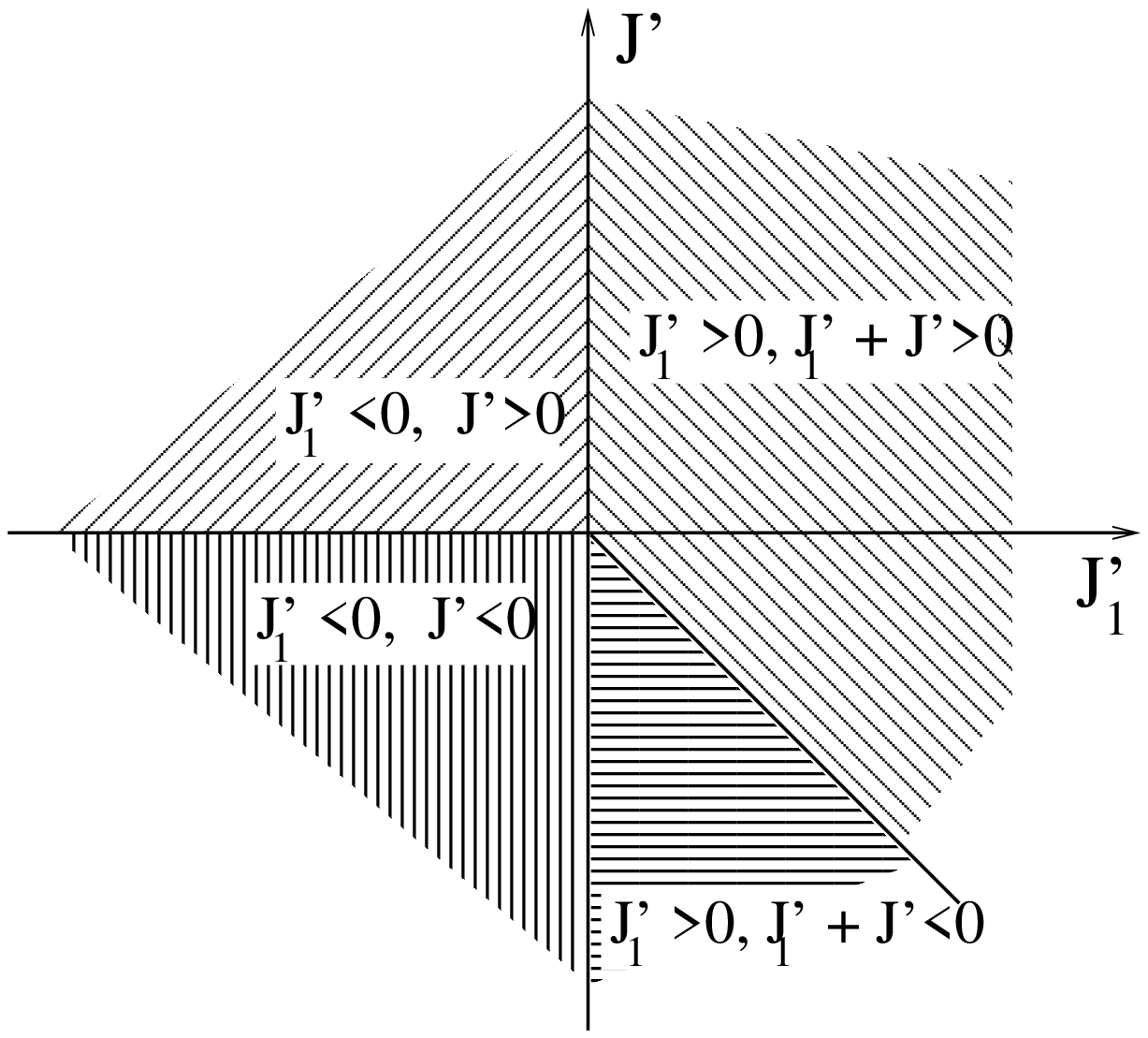}
\end{center}
\caption{Phase diagram of ground states} \label{fig4}
\end{figure}

{\bf Remark 2.} From Corollary \ref{gs5} (see Fig.\ref{fig4}) we
see that when $J\in B_1$ then the model becomes ferromagnetic and
for it there are only three translation-invariant ground states.
When $J\in B_3$ then the model stands antiferromagnetic and hence
it has countable number of periodic ground states. The case $J\in
B_2$ defines dipole ground states. When  $J\in B_4$ then the
ground states determine certain solution of the tricolor problem
on the Bethe lattice. All these results agree with the
experimental ones (see \cite{NS}).

\section{Phase transition}

In this section we are going to describe the existence of a phase
transition for the ferromagnetic Potts model with competing
interactions. We will find a critical curve under one there exists
a phase transition. We also construct the Gibbs measures
corresponding to the ground states $\s^{(i)}$ ($i=1,2,3$) in the
scheme of section 3. Recall that here by a {\it phase transition}
we mean the existence of at least two limiting Gibbs measures (for
more definitions see \cite{Ge},\cite{Pr},\cite{S}).

It should be noted that any transformation $\tau_g$, $g\in \G^2_+$
(see \eqref{trans1}) induces a shift $\tilde\tau_g: \O\to\O$ given
by the formula
\[
(\tilde \tau_g\s)(x)=\s(\tau_g x), \ \  x\in \G^2_+,\ \s\in \O.
\]

A Gibbs measure $\m$ on $\O$ is called {\it translation -
invariant} if for every $g\in \G^2_+$ the equality holds
$\m(\tilde\tau^{-1}_g(A))=\m(A)$ for all $A\in{\mathcal F}$, $g\in
\G^2_+$.

According to section 3 to show the existence of the phase
transition it is enough to find two different solutions of the
equation \eqref{35}, but the analysis of solutions \eqref{35} is
rather tricky. Therefore, it is natural to begin with translation
- invariant ones, i.e. $h_x=h$ is constant for all $x\in V$. Such
kind of solutions will describe translation-invariant Gibbs
measures. In this case the equation \eqref{35} is reduced to the following one
\begin{eqnarray}\label{eq-h}
\left\{
\begin{array}{ll}
u=\frac{\theta_1^2\theta u^2+2\theta_1 uv+\theta v^2+2\theta_1
u+2v+\theta}{\theta u^2+2uv+\theta v^2+2\theta_1 u+2\theta_1 v+\theta_1^2\theta}\\
v=\frac{\theta u^2+2\theta_1 uv+\theta_1^2\theta v^2+2\theta_1
v+2u+\theta}{\theta u^2+2uv+\theta v^2+2\theta_1 u+2\theta_1
v+\theta_1^2\theta},
\end{array}
\right.
\end{eqnarray}
where $u=e^{h_1}$,$v=e^{h_2}$ for a vector $h=(h_1,h_2)$.

Thus for $ \theta_p=1 $ using properties of Markov random fields we get the same
system of equations \eqref{rec0}.

{\bf Remark 3.} From \eqref{eq-h} one can observe that the
equation is invariant with respect to the lines $u=v$, $u=1$ and
$v=1$. It is also invariant with respect to the transformation
$u\to 1/u$, $v\to 1/v$. Therefore, it is enough to consider the
equation on the line $v=1$, since other cases can be reduced to
such a case.

So, rewrite \eqref{eq-h} as follows
\begin{eqnarray}\label{func0}
u=f(u;\theta,\theta_1),
\end{eqnarray}
here
\begin{eqnarray}\label{func2}
f(u;\theta,\theta_1)=\frac{\theta_1^2\theta u^2+4\theta_1
u+2(\theta+1)}{\theta
u^2+2(\theta_1+1)u+\theta_1^2\theta+2\theta_1+\theta}
\end{eqnarray}

From \eqref{func2} we find that \eqref{func0} reduces to the
following
$$
\theta
u^3+(2\theta_1-\theta_1^2\theta+2)u^2+(\theta_1^2\theta+\theta-2\theta_1)u-2(\theta+1)=0
$$
which can be represented by
$$
 (u-1)\bigg(\theta
u^2+(\theta_1+1)(\theta(1-\theta_1)+2)u+2(\theta+1)\bigg)=0.
$$

Thus, $u=1$ is a solution of \eqref{func0}, but to exist a phase
transition we have to find other fixed points of \eqref{func2}. It
means that we have to establish a condition when the following
equation
\begin{eqnarray}\label{kv-t}
\theta u^2+(\theta_1+1)(\theta(1-\theta_1)+2)u+2(\theta+1)=0
\end{eqnarray}
has two positive solutions. Of course, the last one \eqref{kv-t}
has the required solutions if
\begin{eqnarray}\label{con1}
&& (\theta_1+1)(\theta(1-\theta_1)+2)<0,\\ \label{con2}
&&\textrm{the discriminant of \eqref{kv-t} is positive.}
\end{eqnarray}

The condition \eqref{con1} implies that
\begin{eqnarray}\label{con11}
\theta_1>1 \ \ \textrm{and} \ \ \theta>\frac{2}{\theta_1-1}.
\end{eqnarray}

Rewrite the condition \eqref{con2} as follows
\begin{eqnarray}\label{con21}
\bigg((\theta_1^2-1)^2-8\bigg)\theta^2-4\bigg((\theta_1+1)^2(\theta_1-1)+2\bigg)\theta+4(\theta_1+1)^2>0,
\end{eqnarray}
which can be represented by
\begin{eqnarray}\label{con22}
(\theta-\x_1)(\theta-\x_2)>0,
\end{eqnarray} where
\begin{eqnarray}\label{con23}
\x_{1,2}=\frac{2\bigg((\theta_1+1)^2(\theta_1-1)+2\bigg)\mp
4\sqrt{(\theta_1+1)^3+1}}{(\theta_1^2-1)^2-8}. \end{eqnarray}

From \eqref{con21} we obtain that
\begin{eqnarray}\label{con24}
\x_1\cdot\x_2=\frac{4(\theta_1+1)^2}{(\theta_1^2-1)^2-8}.
\end{eqnarray}

Now we are going to compare the condition \eqref{con11} with
solution of \eqref{con22}. To do it, let us consider two cases.

{\tt Case (a)}. Let $(\theta_1^2-1)^2-8>0$. This is equivalent to
$\theta_1>\sqrt{1+2\sqrt{2}}$. Hence,  according to \eqref{con24}
we infer that both $\x_1$ and $\x_2$ are positive. So, the
solution of \eqref{con22} is
\begin{eqnarray}\label{con25}
\theta\in (0,\x_1)\cup (\x_2,\infty).
\end{eqnarray}

From \eqref{con23} we can check that
\[
\x_1<\frac{2}{\theta_1-1}<\x_2.
\]
Therefore, from \eqref{con11},\eqref{con25} we conclude that
$\theta$ should satisfy the following condition
\begin{eqnarray}\label{con112}
\theta>\x_2 \ \ \ \textrm{while} \ \ \theta_1>\sqrt{1+2\sqrt{2}}.
\end{eqnarray}

{\tt Case (b)}. Let $(\theta_1^2-1)^2-8<0$, then this with
\eqref{con11} yields that $1<\theta_1<\sqrt{1+2\sqrt{2}}$. Using
\eqref{con22} and \eqref{con24} one can find that
\begin{eqnarray}\label{con4}
\left\{
\begin{array}{ll}
\theta>\x_1, \ \ \textrm{if} \ \ \theta^*<\theta_1<\sqrt{1+2\sqrt{2}}\\[3mm]
\theta>\frac{2}{\theta_1-1}, \ \ \textrm{if} \ \
1<\theta_1<\theta^*,
\end{array}
\right.
\end{eqnarray}
where  $\theta^*$ is a unique solution of the equation
$(x-1)(\sqrt{(x+1)^3+1}-1)-4=0$\footnote{One can be checked that
the function
\[
g(x)=(x-1)(\sqrt{(x+1)^3+1}-1)
\]
is increasing if $x>1$. Therefore, the equation $g(x)=4$ has a
unique solution $\theta^*$ such that $\theta^*>1$.}.

Consequently, if one of the conditions \eqref{con112} or
\eqref{con4} is satisfied then $f(u,;\theta,\theta_1)$ has three
fixed points $u=1$, $u^*_{1}$ and $u_{2}^*$.

Now we are interested when both $u^*_{1}$ and $u_{2}^*$ solutions
are attractive\footnote{Note that the Jacobian at a fixed point
$(u^*,v^*)$ of \eqref{eq-h} can be calculated as follows
\begin{equation}\label{jac1}
J(u^*,v^*)= \left(
\begin{array}{cc}
\l(u^*,v^*) & \k(u^*,v^*) \\
\k(v^*,u^*) & \l(v^*,u^*)\\
\end{array}
\right),
\end{equation}
here \begin{eqnarray}\label{jac2}
\l(u,v)&=&\frac{2((\theta(\theta_1-u)-(v+\theta_1))u+\theta_1(v+1))}{\theta
u^2+2uv+\theta v^2+2\theta_1 u+2\theta_1
v+\theta_1^2\theta},\\\label{jac3} \k(u,v)&=&\frac{2(1-u)(\theta
v+1+u)}{\theta u^2+2uv+\theta v^2+2\theta_1 u+2\theta_1
v+\theta_1^2\theta}.
\end{eqnarray}}. This occurs when
$$
\frac{d}{du}f(u,;\theta,\theta_1)|_{u=1}>1,
$$ since the function
$f(u,;\theta,\theta_1)$ is increasing and bounded. Hence, a simple
calculation shows that the last condition holds if
\footnote{Indeed, this condition also implies that the eigenvalues
of the Jacobian $J(1,1)$ is less than one (see
\eqref{jac1}-\eqref{jac3}).}
\begin{equation}\label{contr}
 \theta_1>2 \ \ \textrm{and} \ \
\theta>\frac{2}{\theta_1-2}. \end{equation}

If $\theta_1>2$ then the condition \eqref{con4} is not satisfied
since $\theta^*<2$.  Consequently, combining the conditions
\eqref{con112} and \eqref{contr} we establish that if
\begin{equation}\label{pht}
 \theta_1>2 \ \ \textrm{and} \ \
\theta>\max\bigg\{\frac{2}{\theta_1-2},\xi_2\bigg\},
\end{equation}
then $f(u,;\theta,\theta_1)$ has three fixed points, and two of
them $u^*_{1}$ and $u_{2}^*$ are attractive. Without loss of
generality we may assume that $u^*_{1}>u_{2}^*$. Then from
\eqref{kv-t} one sees that
\[
u^*_{1}u_{2}^*=\frac{2(\theta+1)}{\theta}.
\]
which implies that
\begin{equation}\label{infty}
u^*_{1}\to\infty, \ \ u_{2}^*\to 0 \ \ \textrm{as} \ \ \b\to\infty
\end{equation}

Let us denote
\[
h^*_{1,1}=\bigg(\frac{2}{3}\log u^*_{1}, 0\bigg), \ \
h^*_{2,1}=\bigg(\frac{2}{3}\log u^*_{2}, 0\bigg),
\]
which are translation-invariant solutions of \eqref{35}.

According to Remark 2 the vectors
\begin{equation}\label{sol}
\left\{
\begin{array}{ll} h^*_{1,2}=(0,\frac{2}{3}\log u^*_{1}), \ \
h^*_{2,2}=(0,\frac{2}{3}\log u^*_{2})\\[2mm]
h^*_{1,3}=(-\frac{2}{3}\log u^*_{1},-\frac{2}{3}\log u^*_{1}), \ \
h^*_{2,3}=(-\frac{2}{3}\log u^*_{2},-\frac{2}{3}\log u^*_{2})\\
\end{array}
\right. \end{equation} are also translation-invariant solutions of
\eqref{35}. The Gibbs measures corresponding these solutions are
denoted by $\m_{1,i},\m_{2,i}$, $(i=1,2,3$), respectively.

From \eqref{pht} we infer that $(J,J_1)$ belongs to $B_1$.
Furthermore, we assume that  \eqref{pht} is satisfied. This means
in this case there are three ground states for the model.
Therefore, when $\b\to\infty$ certain measures $\m_{1,i},\m_{2,i}$
should tend to the ground states $\{\s^{(1)},\s^{(2)},\s^{(3)}\}$.
Let us choose those ones. Take $\m_{1,1}$, then from \eqref{mes1},
\eqref{22} and \eqref{infty} we have
\begin{eqnarray}\label{beta}
\m_{1,1}(\s(x)=\z_1)&=&\dsf{e^{h^*_{1,1}\z_1}}{e^{h^*_{1,1}\z_1}+e^{h^*_{1,1}\z_2}+e^{h^*_{1,1}\z_3}}\nonumber
\\
&=&\dsf{u^*_1}{u^*_1+2}\to 1 \ \ \textrm{as} \  \b\to\infty,
\end{eqnarray} where $x\in V$.

Similarly, using the same argument we may find
\begin{eqnarray}\label{beta1}
\m_{1,2}(\s(x)=\z_2)\to 1, \ \ \m_{1,3}(\s(x)=\z_3)\to 1 \ \
\textrm{as} \  \b\to\infty.
\end{eqnarray}

Denote these measures by $\m_{k}=\m_{1,k}$, $k=1,2,3$. The
relations \eqref{beta},\eqref{beta1} prompt that the following
should be true
$$
\m_i\to \delta_{\s^{(i)}}\ \ \ \textrm{as} \ \ \b\to\infty,
$$
here $\delta_\s$ is a delta-measure concentrated on $\s$. Indeed,
let us without loss of generality consider the measure $\m_1$. We
know that $\s^{(1)}$ is a ground state, therefore according to
Proposition \ref{gs1} one gets that $H(\s_n|_{V_n})\geq
H(\s^{(1)}|_{V_n})$ for all $\s\in \Omega$ and $n>0$. Hence, it
follows from  \eqref{mes1} that
\begin{eqnarray*}
\m_1(\s^{(1)}|_{V_n})&=&\frac{\exp\{-\b
H(\s^{(1)}|_{V_n})+h^*_{1,1}\z_1|W_n|\}}{\sum\limits_{\tilde\s_n\in\O_{V_n}}
\exp\{-\b H(\tilde\s_n)+h^*_{1,1}\sum\limits_{x\in
W_n}\tilde\s(x)\}}\\
&=&\frac{1}{1+\sum\limits_{\tilde\s_n\in\O_{V_n},\tilde \s_n\neq
\s^{(1)}|_{V_n}}
\frac{\exp\{-\b H(\tilde\s_n)+h^*_{1,1}\sum\limits_{x\in W_n}\tilde\s(x)\}}{\exp\{-\b H(\s^{(1)}|_{V_n})+h^*_{1,1}\z_1|W_n|\}}}\\
&\geq&\frac{1}{1+1/u^*_1}\to 1 \ \ \textrm{as} \ \b\to\infty.
\end{eqnarray*}
The last inequality yields that the required relation.

Consequently, the measures $\m_k$ ($k=1,2,3$) describe pure phases
of the model.

\begin{figure}
\begin{center}
\includegraphics[width=8.07cm]{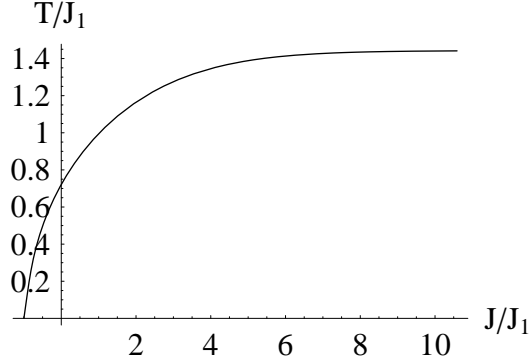}
\end{center}
\caption{The curve
$\frac{T_c}{J_1}=\min\bigg\{\varphi^{-1}\bigg(\frac{J}{J_1}\bigg),\zeta^{-1}\bigg(\frac{J}{J_1}\bigg)\bigg\}$
in the plane $(\frac{J}{J_1},\frac{T}{J_1})$}. \label{fig5}
\end{figure}

Let us find the critical temperature. To do it, rewrite
\eqref{pht} as follows:
\begin{eqnarray}\label{1pht}
\frac{T}{J_1}<\frac{1}{\log 2}, \ \ \
\frac{J}{J_1}>\max\bigg\{\varphi\bigg(\frac{T}{J_1}\bigg),\zeta\bigg(\frac{T}{J_1}\bigg)\bigg\},
\end{eqnarray}
where
\begin{eqnarray*}
\varphi(x)&=&x\log\bigg(\frac{2}{\exp(1/x)-2}\bigg) \\
\zeta(x)&=&x\log\left(\frac{2\bigg((\exp(1/x)+1)^2(\exp(1/x)-1)+2\bigg)+
4\sqrt{(\exp(1/x)+1)^3+1}}{(\exp(2/x)-1)^2-8}\right).
\end{eqnarray*}
From these relations one concludes that the critical line (see
Fig.\ref{fig5})\footnote{Note that the functions $\varphi$ and
$\zeta$ are increasing, therefore their inverse $\varphi^{-1}$ and
$\zeta^{-1}$ exist.} is given by
\begin{eqnarray}\label{crit}
\frac{T_c}{J_1}=\min\bigg\{\varphi^{-1}\bigg(\frac{J}{J_1}\bigg),\zeta^{-1}\bigg(\frac{J}{J_1}\bigg)\bigg\}
\end{eqnarray}

Consequently, we can formulate the following

\begin{thm}\label{ptt}
If the condition \eqref{1pht} is satisfied for the three state
Potts model \eqref{ham1} on the second ordered Bethe lattice, then
there exists a phase transition and  three pure
translation-invariant phases.
\end{thm}

{\bf Remark 4.} If we put $J=0$ to the condition  \eqref{pht} then
the obtained result agrees with the results of \cite{PLM1,PLM2},
\cite{G}.

{\bf Observation.} From \eqref{jac1}-\eqref{jac3} we can derive
that the eigenvalues of the Jacobian at the fixed points
$(u_1^*,1)$, $(1, u_1^*)$, $(u_2^*,1)$, $(1, u_2^*)$,
$((u_1^*)^{-1},(u_1^*)^{-1})$, $((u_2^*)^{-1},(u_2^*)^{-1})$ are
real. Therefore, in this case (i.e. $J_p=0$), there are not the
modulated phases and Lifshitz points. On the other hand, the
absolute value of the eigenvalues of the Jacobian at the fixed
points $(u_1^*,1)$, $(1, u_1^*)$ and $((u_1^*)^{-1},(u_1^*)^{-1})$
are smaller than 1. The absolute value of the eigenvalues at the
fixed points $(u_2^*,1)$, $(1, u_2^*)$ and
$((u_2^*)^{-1},(u_2^*)^{-1})$ are bigger than 1. These show that
the points $(u_1^*,1)$, $(1, u_1^*)$ and
$((u_1^*)^{-1},(u_1^*)^{-1})$ are the stable fixed points of the
transformation given by \eqref{eq-h}. The Gibbs measures
associated with these points are pure phases.

{\bf Remark 5.} Recall that the a Gibbs measure $\m_0$
corresponding to the solution $h=(0,0)$ is called unordered phase.
The purity of the unordered phase was investigated in
\cite{GR},\cite{MR3} when $J=0$. Such a property relates to the
reconstruction thresholds and percolation on lattices (see
\cite{Mar},\cite{JM}). For $J\neq 0$ the purity of $\m_0$ is an
open problem.

\section{ A formula of the free energy}

This section is devoted to the free energy and exact calculation
of certain physical quantities. Since the Bethe lattice is
non-amenable, so we have to prove the existence of the free
energy.

Consider the partition function $Z^{(n)}(\b, h)$ (see
\eqref{partition}) of the Gibbs measure $\m^h_\b$ (which
corresponds to solution $h=\{h_x, x\in V\}$ of the equation
\eqref{35})
\[
Z^{(n)}(\b,h)=\sum_{\tilde\s_n\in\Omega_{V_n}}\exp \{-\b
H(\tilde\s_n)+\sum_{x\in W_n}h_x\tilde\s_n(x)\}.
\]
The free energy is defined by
\begin{eqnarray}\label{free}
 F_\b(h)=-\lim_{n\to \infty}{1\over
3\b\cdot 2^n} \ln Z^{(n)}(\b, h).
\end{eqnarray}
The goal of this section is to prove following:

\begin{thm}\label{free1} The free energy of the model \eqref{ham1} exists for all $h$, and is given by the
formula
\begin{eqnarray}\label{free2}
F_\b(h)=-\lim_{n\to\infty}{1\over 3\b\cdot 2^n}
\sum_{k=0}^n\sum_{x\in W_{n-k}}\log
a(x,h_y,h_z;\theta,\theta_1,\b),
\end{eqnarray} where $y=y(x),
z=z(x)$ are direct successors of $x$;
\begin{eqnarray}\label{ax}
a(x,h_y,h_z;\theta,\theta_1,\b)=e^{-(J/2+J_1)\b}g(h'_y,h'_z)\bigg[F(h'_y,h'_z)F((h'_y)^t,(h'_z)^t)\bigg]^{1/3},
\end{eqnarray}
here the function $F(h,r)$ is defined as in \eqref{func1}, and
\[
g(h,r)=\theta e^{h_1+r_1}+e^{h_1+r_2}+ e^{h_2+r_1}+\theta
e^{h_2+r_2}+\theta_1(e^{h_1}+e^{r_1}+e^{h_2}+e^{r_2})+\theta_1^2\theta,
\]
where $h=(h_1,h_2), r=(r_1,r_2)$.
\end{thm}

\begin{proof} We shall use the recursive equation \eqref{An}, i.e.
\[
Z^{(n)}=A_{n-1}Z^{(n-1)},
\]
where $A_n=\prod\limits_{x\in W_n}a(x,h_y,h_z;\theta,\theta_1,\b)$
 $x\in V$, $y,z\in S(x)$, which is defined below. Using \eqref{eq5} we
have \eqref{ax}.

Thus, the recursive equation \eqref{An} has the following form
\begin{eqnarray}\label{free3}
Z^{(n)}(\b;h)=\exp\bigg(\sum_{x\in W_{n-1}}\log
a(x,h_y,h_z;\theta,\theta_1,\b)\bigg) Z^{(n-1)}(\b,h).
\end{eqnarray}

Now we prove existence of the RHS limit of \eqref{free2}. From the
form of the function $F$ one gets that it is bounded, i.e.
$|F(h,r)|\leq M$ for all $h,r\in\br^2$. Hence,  we conclude that
the solutions of the equation \eqref{35} are bounded, i.e.
$|h_{x,i}|\leq C$ for all $x\in V$, $i=1,2$. Here $C$ is some
constant and $h_x=(h_{x,1}h_{x,2})$. Consequently the function
$a(x,h_y,h_z;\theta,\theta_1,\b)$ is bounded, and so $|\log
a(x,h_y,h_z;\theta,\theta_1,\b)|\leq C_\b$ for all $h_y, h_z$.
Hence we get
\begin{eqnarray}\label{45}
&&{1\over 3\cdot 2^n}\sum_{k=l+1}^n\sum_{x\in W_{n-k}}\log
a(x,h_y,h_z;\theta,\theta_1,\b) \nonumber\\
&&\leq{C_\b\over  2^n}\sum_{k=l+1}^n 2^{n-k-1} \leq C_\b\cdot
2^{-l}. \end{eqnarray}
 Therefore, from \eqref{45} we get the existence of the limit at
RHS of \eqref{free2}.
\end{proof}

Let us compute the free energy corresponding the measures $\m_i$,
($i=1,2,3$). Assuming first that $h_x=h$ for all $x\in V$. Then
from \eqref{free2} and \eqref{ax} one gets
$$F_\b(h)=\frac{1}{\b}\log a(h,\theta,\theta_1,\b),$$
here
\begin{eqnarray}\label{ax1}
a(h,\theta,\theta_1,\b)=e^{-(J/2+J_1)\b}g(h',h')\bigg[F(h',h')F((h')^t,(h')^t)\bigg]^{1/3}.
\end{eqnarray}

Let us consider $h=h^*_{1,k}$, ($k=1,2,3$). Denote
$F_{\b}=F_\b(h^*_{1,k})$. Then we have
\begin{eqnarray}\label{free4}
\b F_{\b}=\log\bigg[e^{-(J/2+J_1)\b}(u^*_1)^{1/3}(\theta
(u^*_1)^2+2(\theta_1+1)u^*_1+\theta_1^2\theta+2\theta_1+\theta)\bigg].
\end{eqnarray}

Taking into account \eqref{kv-t} the equality \eqref{free4} can be
rewritten as follows:
\begin{eqnarray}\label{free4}
\b F_{\b}=-(J/2+J_1)\b+\frac{1}{3}\log u^*_1+\log(\theta_1-1)+
\log\bigg[\theta(\theta_1+1)(u^*_1+1)+2\bigg].
\end{eqnarray}

Now let us compute the internal energy $U$ of the model. It is
known that the following formula holds
\begin{equation}\label{ener}
U=\frac{\partial(\b F_{\b})}{\partial\b}.
\end{equation}

Before compute it we have to calculate $du^*_1/d\b$. Taking
derivation from both sides of \eqref{kv-t} one finds
\begin{equation}\label{der}
\frac{du^*_1}{d\b}=\frac{3\bigg((J_1\theta_1(\theta\theta_1-1)+J(\theta_1+1))u^*_1+J\bigg)}{2\theta
u^*_1+(\theta_1+1)(\theta(1-\theta_1)+2)}.
\end{equation}

From \eqref{free4} and \eqref{ener} we obtain
\begin{eqnarray*}
U&=&-(J/2+J_1)+\frac{3}{2}\bigg[
\frac{J_1}{\theta_1-1}+\frac{\theta((J+J_1)\theta_1+J)(u^*_1+1)}{\theta(\theta_1+1)(u^*_1+1)+2}\bigg]\\
&&+\bigg[\frac{\theta(\theta_1+1)(4u^*_1+1)+2}{3u^*_1(\theta(\theta_1+1)(u^*_1+1)+2)}\bigg]\frac{du^*_1}{d\b}
\end{eqnarray*}

Again using \eqref{kv-t} and \eqref{der} one gets
\begin{eqnarray*}
U&=&-(J/2+J_1)+\frac{3}{2}\bigg[
\frac{\theta\theta_1(J_1(\theta^2_1+1)+J(\theta_1-1))(u^*_1+1)+2J_1}{\theta(\theta^2_1-1)(u^*_1+1)+2(\theta_1-1)}\bigg]\\
&&+\bigg[\frac{\theta(\theta_1+1)(4u^*_1+1)+2}{(\theta(\theta_1+1)(\theta\theta_1-2)-2)u^*_1-2(\theta+1)(\theta_1+1)}\bigg]\\
&&\times\bigg[\frac{((J_1\theta_1(\theta\theta_1-1)+J(\theta_1+1))u^*_1+J)}{2\theta
u^*_1+(\theta_1+1)(\theta(1-\theta_1)+2)}\bigg]
\end{eqnarray*}

Using this expression we can also calculate entropy of the model.

Since spins take values in $\br^2$, therefore the magnetization of
the model would be $\br^2$-valued quantity. Using the result of
sections 4 and 5 we can easily compute the magnetization. Let us
calculate it with respect to the measure $\m_1$. Note that the
model is translation-invariant, therefore, we have $
M_1=<\s_{(0)}>_{\m_1}$, so using \eqref{22}, \eqref{zero} and
\eqref{mes1} one finds
\begin{eqnarray*}
M_1&=&\z_1\m_1(\s_{(0)}=\z_1)+\z_2\m_1(\s_{(0)}=\z_2)+\z_3\m_1(\s_{(0)}=\z_3)\nonumber\\
&=&\frac{1}{(u_1^*)^{2/3}+2(u_1^*)^{-1/3}}\bigg(\z_1(u_1^*)^{2/3}+\z_2(u_1^*)^{-1/3}+\z_3(u_1^*)^{-1/3}\bigg)\nonumber\\
&=&\frac{1}{u_1^*+2}\bigg(\z_1u_1^*+\z_2+\z_3\bigg)\nonumber\\
&=&\frac{u_1^*-1}{u_1^*+2}\z_1.\nonumber
\end{eqnarray*}

Similarly, one gets
\begin{eqnarray*}
M_2=<\s_{(0)}>_{\m_2}&=&\frac{u_1^*-1}{u_1^*+2}\z_2.\nonumber \\
M_3=<\s_{(0)}>_{\m_3}&=&\frac{u_1^*-1}{2u_1^*+1}\z_3.\nonumber
\end{eqnarray*}

\section{Discussion of results}

It is known \cite{Ba} that to exact calculations in statistical
mechanics are paid attention by many of researchers, because those
are important not only for their own interest but also for some
deeper understanding of the critical properties of spin systems
which are not obtained form approximations. So, those are very
useful for testing the credibility and efficiency of any new
method or approximation before it is applied to more complicated
spin systems. In the present paper we have derived recurrent
equations for the partition functions of the three state Potts
model with competing interactions on a Bethe lattice of order two,
and certain particular cases of those equations were studied. In
the presence of the one-level competing interactions we exactly
solved the ferromagnetic Potts model. The critical curve
\eqref{crit} such that there exits a phase transitions under it,
was calculated (see Fig. \ref{fig5}). It has been described the
set of ground states of the model (see Fig. \ref{fig4}). This
shows that the ground states of the model are richer than the
ordinary Potts model on the Bethe lattice. Using this description
and the recurrent equations, one found the Gibbs measures
associated with the translation-invariant ground states. Note that
such Gibbs measures determine generalized 2-step Markov chains
(see \cite{D}). Moreover, we proved the existence of the free
energy, and exactly calculated it for those measures. Besides, we
have computed some other physical quantities too. The results
agrees with \cite{PLM1,PLM2}, \cite{G} when we neglect the next
nearest neighbor interactions.

Note that for the Ising model on the Bethe lattice with in the
presence of the one-level and prolonged competing interactions the
modulated phases and Lifshitz points appear in the phase diagram
(see \cite{V},\cite{YOS},\cite{SC}). In absence of the prolonged
competing interactions in the 3-state Potts model we do not have
such kind of phases, this means one-level interactions could not
affect the appearance the modulated phases. One can hope that the
considered Potts model with $J_p=0$ will describe some biological
models. Note that the case, when the prolonged competing
interaction is nontrivial ($J_p\neq 0$),  will be a theme of our
next investigations \cite{GMMP}, where the modulated phases and
Lifshitz points will be discussed.

\section*{Acknowledgements} F.M. thanks the FCT (Portugal) grant SFRH/BPD/17419/2004.
J.F.F.M. acknowledges projects POCTI/FAT/46241/2002,
POCTI/MAT/46176/2002 and European research NEST project DYSONET/
012911.The work was also partially supported by Grants:
$\Phi$-1.1.2, $\Phi$-2.1.56 of CST of the Republic of Uzbekistan.

\appendix

\section{Recurrent equations at $J_p\neq 0$}

Denote
\begin{eqnarray}\label{ABC}
\left\{
\begin{array}{lll}
A^{(n)}_1&=&\theta^2_p Z^{(n)}_1+\theta_p Z^{(n)}_2+\theta_p Z^{(n)}_3+\theta_p Z^{(n)}_4+ Z^{(n)}_5+ Z^{(n)}_6+ \theta_p Z^{(n)}_7+ Z^{(n)}_8+ Z^{(n)}_9, \\
B^{(n)}_1&=&\theta^2_p Z^{(n)}_{10}+\theta_p Z^{(n)}_{11}+\theta_p Z^{(n)}_{12}+\theta_p Z^{(n)}_{13}+ Z^{(n)}_{14}+ Z^{(n)}_{15}+ \theta_p Z^{(n)}_{16}+ Z^{(n)}_{17}+ Z^{(n)}_{18} ,\\
C^{(n)}_1&=&\theta^2_p Z^{(n)}_{19}+\theta_p Z^{(n)}_{20}+\theta_p Z^{(n)}_{21}+\theta_p Z^{(n)}_{22}+ Z^{(n)}_{23}+ Z^{(n)}_{24}+ \theta_p Z^{(n)}_{25}+ Z^{(n)}_{26}+ Z^{(n)}_{27} ,\\
A^{(n)}_2&=& Z^{(n)}_1+\theta_p Z^{(n)}_2+ Z^{(n)}_3+\theta_p Z^{(n)}_4+ \theta_p^2 Z^{(n)}_5+ \theta_p Z^{(n)}_6+  Z^{(n)}_7+ \theta_p Z^{(n)}_8+ Z^{(n)}_9, \\
B^{(n)}_2&=& Z^{(n)}_{10}+\theta_p Z^{(n)}_{11}+ Z^{(n)}_{12}+\theta_p Z^{(n)}_{13}+ \theta_p^2 Z^{(n)}_{14}+ \theta_p Z^{(n)}_{15}+  Z^{(n)}_{16}+ \theta_p Z^{(n)}_{17}+ Z^{(n)}_{18}, \\
C^{(n)}_2&=& Z^{(n)}_{19}+\theta_p Z^{(n)}_{20}+ Z^{(n)}_{21}+\theta_p Z^{(n)}_{22}+ \theta_p^2 Z^{(n)}_{23}+ \theta_p Z^{(n)}_{24}+  Z^{(n)}_{25}+ \theta_p Z^{(n)}_{26}+ Z^{(n)}_{27}, \\
A^{(n)}_3&=& Z^{(n)}_1+Z^{(n)}_2+ \theta_p Z^{(n)}_3+ Z^{(n)}_4+  Z^{(n)}_5+ \theta_p Z^{(n)}_6+ \theta_p Z^{(n)}_7+ \theta_p Z^{(n)}_8+ \theta_p^2 Z^{(n)}_9, \\
B^{(n)}_3&=& Z^{(n)}_{10}+Z^{(n)}_{11}+ \theta_p Z^{(n)}_{12}+ Z^{(n)}_{13}+  Z^{(n)}_{14}+ \theta_p Z^{(n)}_{15}+ \theta_p Z^{(n)}_{16}+ \theta_p Z^{(n)}_{17}+ \theta_p^2 Z^{(n)}_{18}, \\
C^{(n)}_3&=& Z^{(n)}_{19}+Z^{(n)}_{20}+ \theta_p Z^{(n)}_{21}+
Z^{(n)}_{22}+  Z^{(n)}_{23}+ \theta_p Z^{(n)}_{24}+ \theta_p
Z^{(n)}_{25}+ \theta_p Z^{(n)}_{26}+ \theta_p^2 Z^{(n)}_{27},
\end{array}
\right.
\end{eqnarray}
then the last one in terms of \eqref{den} is represented by
\begin{eqnarray}\label{ABC1}
\left\{
\begin{array}{lll}
A^{(n)}_{1}&=& \theta_p^2 x^{(n)}_1+2\theta_p x^{(n)}_2+2\theta_p x^{(n)}_3+x^{(n)}_4+2x^{(n)}_5+x^{(n)}_6 ,\\
B^{(n)}_{1}&=& \theta_p^2 x^{(n)}_7+2\theta_p x^{(n)}_8+2\theta_p x^{(n)}_9+x^{(n)}_{10}+2x^{(n)}_{11}+x^{(n)}_{12} ,\\
C^{(n)}_{1}&=& \theta_p^2 x^{(n)}_{13}+2\theta_p x^{(n)}_{14}+2\theta_p x^{(n)}_{15}+x^{(n)}_{16}+2x^{(n)}_{17}+x^{(n)}_{18} ,\\
A^{(n)}_{2}&=& x^{(n)}_1 +2\theta_p x^{(n)}_2+2 x^{(n)}_3+\theta^2_p x^{(n)}_4+2 x^{(n)}_5+x^{(n)}_6 ,\\
B^{(n)}_{2}&=& x^{(n)}_7+2\theta_p x^{(n)}_8+2 x^{(n)}_9+ \theta^2_p x^{(n)}_{10}+2x^{(n)}_{11}+x^{(n)}_{12} ,\\
C^{(n)}_{2}&=&  x^{(n)}_{13}+2\theta_p x^{(n)}_{14}+2 x^{(n)}_{15}+\theta^2_px^{(n)}_{16}+2x^{(n)}_{17}+x^{(n)}_{18} ,\\
A^{(n)}_{3}&=& x^{(n)}_1+2 x^{(n)}_2+2\theta_p x^{(n)}_3+x^{(n)}_4+2\theta_p x^{(n)}_5 + \theta_p^2 x^{(n)}_6 ,\\
B^{(n)}_{3}&=& x^{(n)}_7+2 x^{(n)}_8+2\theta_p x^{(n)}_9+x^{(n)}_{10}+2\theta_p x^{(n)}_{11}+\theta_p^2 x^{(n)}_{12} ,\\
C^{(n)}_{3}&=&  x^{(n)}_{13}+ 2x^{(n)}_{14}+2\theta_p
x^{(n)}_{15}+x^{(n)}_{16}+2\theta_px^{(n)}_{17}+\theta_p^2
x^{(n)}_{18}.
\end{array}
\right.
\end{eqnarray}

From \eqref{tZ},\eqref{ABC} and \eqref{ABC1} we obtain
\begin{eqnarray*}
A^{(n)}_{1}&=& \tilde Z^{(n)}_{1} + (\theta_p^2 -1) x^{(n)}_1+2(\theta_p -1) x^{(n)}_2+2(\theta_p -1) x^{(n)}_3 ,\\
B^{(n)}_{1}&=& \tilde Z^{(n)}_{2} + (\theta_p^2 -1) x^{(n)}_7+2(\theta_p -1) x^{(n)}_8+2(\theta_p -1) x^{(n)}_9 ,\\
C^{(n)}_{1}&=& \tilde Z^{(n)}_{3} + (\theta_p^2 -1) x^{(n)}_{13}+2(\theta_p -1) x^{(n)}_{14}+2(\theta_p -1) x^{(n)}_{15} ,\\
A^{(n)}_{2}&=& \tilde Z^{(n)}_{1} + 2(\theta_p -1)x^{(n)}_2 +(\theta_p^2 -1) x^{(n)}_4+2(\theta_p -1) x^{(n)}_5 ,\\
B^{(n)}_{2}&=& \tilde Z^{(n)}_{2} + 2(\theta_p -1)x^{(n)}_8 +(\theta_p^2 -1) x^{(n)}_{10}+2(\theta_p -1) x^{(n)}_{11} ,\\
C^{(n)}_{2}&=& \tilde Z^{(n)}_{3} + 2(\theta_p -1)x^{(n)}_{14} +(\theta_p^2 -1) x^{(n)}_{16}+2(\theta_p -1) x^{(n)}_{17} ,\\
A^{(n)}_{3}&=& \tilde Z^{(n)}_{1} + 2(\theta_p -1)x^{(n)}_3 +2(\theta_p -1) x^{(n)}_5+(\theta_p^2 -1) x^{(n)}_6 ,\\
B^{(n)}_{3}&=& \tilde Z^{(n)}_{2} + 2(\theta_p -1)x^{(n)}_9
+2(\theta_p
-1) x^{(n)}_{11}+(\theta_p^2 -1) x^{(n)}_{12} ,\\
C^{(n)}_{3}&=& \tilde Z^{(n)}_{3} + 2(\theta_p -1)x^{(n)}_{15}
+2(\theta_p -1) x^{(n)}_{17}+(\theta_p^2 -1) x^{(n)}_{18} .
\end{eqnarray*}

Now let us assume that $J_p\neq 0$ and $\bar\s\equiv\z_1$. Then
\begin{eqnarray*}
&&B_1^{(n)}=C_1^{(n)}, \quad A_2^{(n)}=A_3^{(n)},\\
&&B_2^{(n)}=C_3^{(n)}, \quad B_3^{(n)}=C_2^{(n)},
\end{eqnarray*}
and
$$
\tilde Z_2^{(n)}= \tilde Z_3^{(n)}.
$$
Hence the recurrence system \eqref{rec11} has the following form
\begin{eqnarray}\label{rec12}
\left\{
\begin{array}{ll}
x^{(n+1)}_1= \theta\theta_1^2 (A^{(n)}_1)^2, &
x^{(n+1)}_2= x^{(n+1)}_3=\theta_1 A^{(n)}_1B^{(n)_1}, \\
x^{(n+1)}_4= x^{(n+1)}_6 =\theta (B^{(n)}_1)^2,& x^{(n+1)}_5 =
(B^{(n)}_1)^2 ,\\
x^{(n+1)}_7 = \theta (A^{(n)}_2)^2, &
x^{(n+1)}_8 =  \theta_1A^{(n)}_2 B^{(n)}_2,  \\
x^{(n+1)}_9 =  A^{(n)}_2 C^{(n)}_2, &
x^{(n+1)}_{10} = \theta\theta_1^2  (B^{(n)}_2)^2,  \\
x^{(n+1)}_{11} = \theta_1  B^{(n)}_2 C^{(n)}_2 , & x^{(n+1)}_{12}
= \theta (C_2^{(n)})^2,\\
x^{(n+1)}_{13}= x^{(n+1)}_{7}, & x^{(n+1)}_{14}=x^{(n+1)}_{9},\\
x^{(n+1)}_{15}=x^{(n+1)}_{8}, & x^{(n+1)}_{16}=x^{(n+1)}_{12},\\
x^{(n+1)}_{17}=x^{(n+1)}_{11}, & x^{(n+1)}_{18}=x^{(n+1)}_{10}.
\end{array}
\right.
\end{eqnarray}

Through introducing new variables
\begin{eqnarray*}
\left\{
\begin{array}{ll}
y^{(n)}_1=x^{(n)}_1, & y^{(n)}_2=x^{(n)}_2=x^{(n)}_3,\\
y^{(n)}_3=x^{(n)}_5=\frac{x^{(n)}_4}{\theta}=\frac{x^{(n)}_6}{\theta},
&\\
y^{(n)}_4=x^{(n)}_{7}=x^{(n)}_{13} ,&
y^{(n)}_5=x^{(n)}_{8}=x^{(n)}_{15},\\
y^{(n)}_6=x^{(n)}_{9}=x^{(n)}_{14} , &
y^{(n)}_7=x^{(n)}_{10}=x^{(n)}_{18},\\
y^{(n)}_8=x^{(n)}_{11}=x^{(n)}_{17}, &
y^{(n)}_9=x^{(n)}_{12}=x^{(n)}_{16},
\end{array}
\right.
\end{eqnarray*}
the recurrence system \eqref{rec12} takes the following form
\begin{eqnarray*}
\left\{
\begin{array}{ll}
y^{(n+1)}_1= \theta\theta_1^2 (\tilde A^{(n)}_1)^2 ,&
y^{(n+1)}_2=\theta_1 \tilde A^{(n)}_1\tilde B^{(n)}_1, \\
y^{(n+1)}_3= (\tilde B^{(n)}_1)^2,& y^{(n+1)}_4 = \theta (\tilde
A^{(n)}_2)^2, \\
y^{(n+1)}_5 =  \theta_1\tilde A^{(n)}_2 \tilde B^{(n)}_2,  &
y^{(n+1)}_6 =  \tilde A^{(n)}_2\tilde C^{(n)}_2,\\
y^{(n+1)}_{7} = \theta\theta_1^2  (\tilde B^{(n)}_2)^2,  &
y^{(n+1)}_{8} = \theta_1 \tilde B^{(n)}_2 \tilde C^{(n)}_2 , \\
y^{(n+1)}_{9}= \theta (\tilde C_2^{(n)})^2, &
\end{array}
\right.
\end{eqnarray*}
where
\begin{eqnarray*}
\tilde A^{(n)}_{1}&=& \theta_p^2 y^{(n)}_1+4\theta_p y^{(n)}_2+2(\theta+1) y^{(n)}_3 ,\\
\tilde B^{(n)}_{1}&=& \theta_p^2 y^{(n)}_4+2\theta_p y^{(n)}_5+2\theta_p y^{(n)}_6+ y^{(n)}_7+2y^{(n)}_8+y^{(n)}_9,\\
\tilde A^{(n)}_{2}&=& y^{(n)}_1+ (2\theta_p +1) y^{(n)}_{8}+(\theta_p^2\theta+2\theta_p+\theta) y^{(n)}_{3} ,\\
\tilde B^{(n)}_{2}&=& y^{(n)}_4 + 2\theta_p y^{(n)}_5 +2y^{(n)}_6+\theta_p^2 y^{(n)}_{7}+2\theta_p y^{(n)}_{8}+y^{(n)}_9 ,\\
\tilde C^{(n)}_{2}&=& y^{(n)}_4 +2\theta_p
y^{(n)}_6+2y^{(n)}_5+\theta_p^2 y^{(n)}_{9}+2\theta_p
y^{(n)}_{8}+y^{(n)}_7 .
\end{eqnarray*}

Noting that
\begin{eqnarray*}
\begin{array}{ll}
\theta( y^{(n)}_2)^2= y^{(n)}_1 y^{(n)}_3, & \theta^2( y^{(n)}_5)^2= y^{(n)}_4 y^{(n)}_7,\\
\theta^2( y^{(n)}_6)^2= y^{(n)}_4 y^{(n)}_9, & \theta^2(
y^{(n)}_8)^2= y^{(n)}_7 y^{(n)}_9 ,
\end{array}
\end{eqnarray*}
we see that only five independent variables remain.

It should be noted that if $\theta_1=1$, i.e. $J_1=0$, then for
the boundary condition $\bar\s\equiv\z_1$ we have
\begin{eqnarray*}
&&A_1^{(n)}=A_2^{(n)}=A_3^{(n)}=\theta_p^4 B_1^{(n)},\\
&& B_1^{(n)}=B_2^{(n)}=B_3^{(n)}=C_1^{(n)}=C_2^{(n)}=C_3^{(n)},\\
&&\tilde Z_2^{(n)}= \tilde Z_3^{(n)},
\end{eqnarray*}
so that
\begin{eqnarray*}
\frac{\tilde Z_1^{(n+1)}}{\tilde Z_3^{(n+1)}}=
\frac{\theta(\theta_p^4B_1^{(n)})^2+4(\theta_p^4B_1^{(n)})B_1^{(n)}+2(\theta+1)(B_1^{(n)})^2}
{\theta(\theta_p^4B_1^{(n)})^2+4(\theta_p^4B_1^{(n)})B_1^{(n)}+2(\theta+1)(B_1^{(n)})^2}=1.
\end{eqnarray*}
Consequently, when $\theta_1=1 $ for any boundary condition exists
single limit Gibbs measure, namely, the unordered phase. So that
the phase transition does not occur.

\section{Proof of the consistency condition}

In this section we show that the condition \eqref{consis1} and
\eqref{35} are equivalent. Assume that \eqref{consis1} holds. Then
inserting \eqref{mes1} into \eqref{consis1} we find
\begin{eqnarray}\label{eq3}
&&\frac{Z^{(n-1)}}{Z^{(n)}}\prod_{x\in
W_{n-1}}\sum_{\sigma_x^{(n)}}
\exp\{\beta J_1\s(x)(\s(y)+\s(z))+\beta J\s(y)\s(z) \nonumber\\
&&+ h_y\sigma(y)+h_z\s(z)\}= \prod_{x \in
W_{n-1}}\exp\{h_x\sigma(x)\}, \end{eqnarray} here given $ x\in
W_{n-1}$ we denoted $S(x)=\{y,z\},\ \ \sigma_x^{(n)}=
\{\sigma(y),\s(z)\}$ and used $ \sigma^{(n)}=\bigcup\limits_{x\in
W_{n-1}} \sigma_x^{(n)}. $

Now fix $x\in W_{n-1}$ and rewrite \eqref{eq3} for the cases $
\sigma(x)=\z_i $ ($i=1,2$) and $ \sigma(x)=\z_3 $, and then
taking their rations we find
\begin{eqnarray}\label{eq4}
&&{\sum_{\sigma_x^{(n)}=\{\s(y),\s(z)\}} \exp\{\beta
J_1\z_i(\s(y)+\s(z))+\beta J\s(y)\s(z)+h_y\sigma(y)+h_z\s(z)\}
\over \sum_{\sigma_x^{(n)}=\{\s(y),\s(z)\}} \exp\{-\beta
J_1\z_3(\s(y)+\s(z))+\beta J\s(y)\s(z)+
h_y\sigma(y)+h_z\s(z)\}}\nonumber\\
&& =\exp\{h'_{x,i}\}.
\end{eqnarray}

Now by using \eqref{22} from \eqref{eq4} we get
\begin{eqnarray}\label{eq5}
e^{h_{x,1}'}=F(h_y',h_z'), \ \ \
e^{h_{x,2}'}=F((h_y')^t,(h_z')^t).
\end{eqnarray}

From the equality  \eqref{eq5} we conclude that the function
${\mathbf{h}}=\{h_x=(h_{x,1},h_{x,2}): x\in V\}$ should satisfy
\eqref{35}.

Note that the converse is also true, i.e. if \eqref{35} holds that
measures defined by \eqref{mes1} satisfy the consistency
condition. Indeed, the equality \eqref{35} implies \eqref{eq5},
and hence \eqref{eq4}. From \eqref{eq4} we obtain
\begin{eqnarray*}
&&\sum_{\sigma_x^{(n)}=\{\s(y),\s(z)\}} \exp\{\beta
J_1\z_i(\s(y)+\s(z))+\beta J\s(y)\s(z)+h_y\sigma(y)+h_z\s(z)\}\\
&&= a(x)\exp\{\z_ih_x\}, \end{eqnarray*} where $i=1,2,3$ and
$a(x)$ is some function. This equality implies
\begin{eqnarray}\label{eq6}
&&\prod_{x\in W_{n-1}} \sum_{\sigma_x^{(n)}=\{\s(y),\s(z)\}}
\exp\{\beta J_1(\s(y)+\s(z))\s(x)+\beta
J\s(y)\s(z)+h_y\sigma(y)+h_z\s(z)\}\nonumber\\
&&=\prod_{x\in W_{n-1}}a(x)\exp\{\s(x)h_x\}. \end{eqnarray}
 Writing
$A_n=\prod_{x\in W_n}a(x) $ from \eqref{eq6} one gets
\begin{eqnarray}\label{eq61}
Z^{(n-1)}A_{n-1}\mu^{(n-1)}(\sigma_{n-1})=
Z^{(n)}\sum_{\sigma^{(n)}}\mu_n(\sigma_{n-1}\vee\sigma^{(n)}).
\end{eqnarray} Taking into account that each $\mu^{(n)},\ \  n\geq 1$ is a probability
measure, i.e.
$$
\sum_{\s_{n-1}}\sum_{\s^{(n)}}\mu^{(n)}(\s_{n-1}\vee \s^{(n)})=1,
\ \ \sum_{\s_{n-1}}\mu^{(n-1)}(\s_{n-1})=1,$$ from \eqref{eq61} we
infer
\begin{eqnarray}\label{An}
Z^{(n-1)}A_{n-1}=Z^{(n)}, \end{eqnarray}
 which means that \eqref{consis1} holds.

\end{document}